\documentclass[12pt]{article}
\usepackage{epsfig} \usepackage{changebar}
\def\beq{\begin{equation}}
\def\eeq{\end{equation}}
\def\bea{\begin{eqnarray}}
\def\eea{\end{eqnarray}}
\def\bq{\begin{quote}}
\def\eq{\end{quote}}

\def\beqa{\begin{eqnarray}} 
\def\eeqa{\end{eqnarray}} 
\def\be{\begin{equation}} 
\def\ee{\end{equation}} 
\def\beq{\begin{equation}}   
\def\eeq{\end{equation}}

\def\bi{\begin{itemize}} 
\def\ei{\end{itemize}} 
 
\def\eps{\epsilon}

\def\gappeq{\mathrel{\rlap
{\raise.5ex\hbox{$>$}}
{\lower.5ex\hbox{$\sim$}}}}
\def\lappeq{\mathrel{\rlap{\raise.5ex\hbox{$<$}}
{\lower.5ex\hbox{$\sim$}}}}
\begin{document}
\pagestyle{empty}
\begin{flushright}
BONN-TH-99-12\\
{hep-th/9908147}
\end{flushright}
\vspace{1.0cm}
\begin{center}
{\bf \Large Scales of Gaugino Condensation 
and Supersymmetry Breaking in Nonstandard M-Theory Embeddings}\\
\vspace*{2.0cm} 
Zygmunt Lalak$^{a,b)}$  and  Steven Thomas $^{c)}$ 
{ }\\
\vspace{0.3cm}
\vspace*{1.0cm}  
{\bf ABSTRACT } \\
\begin{flushleft}
We investigate the formation of dynamical gaugino condensates  
and supersymmetry breaking 
in the compactifications of  Horava-Witten theory with perturbative 
nonstandard embeddings.
Specific models are considered where the underlying massless charged 
states of the condensing sector  
are determined by the spectra of  $Z_2 \times  Z_2 $ and $Z_4 $ 
orbifolds with nonstandard embeddings. We find among them viable examples 
where gaugino condensation is triggered on the wall with 
the weakest gauge coupling at $M_{GUT} $. In all these cases the magnitude 
of the condensate formed is below the energy scales at which extra dimensions 
are resolved, and so justifies the analysis of condensation in an effective 
4-dimensional framework. We make some comments concerning  the size of the
 largest extra dimension in the models considered. We discuss racetrack 
scenarios in the framework of perturbative nonstandard embeddings.
\end{flushleft}
\end{center}
\vspace*{5mm}      
\noindent
{\bf }
 
\vspace*{1.0cm} 
\noindent

\rule[.1in]{16.5cm}{.002in}

\noindent
$^{a)}$ Physikalishes Institut, Universitat Bonn, Germany\\
$^{b)}$ Institute of Theoretical Physics, Warsaw University, Poland\\
$^{c)}$ Department of Physics, Queen Mary and Westfield College, London UK \\
\vspace*{0.5cm}
\begin{flushleft} 
August 1999
\end{flushleft}
\vfill\eject

\setcounter{page}{1}
\pagestyle{plain}

\section{Introduction}
Recent advances in our understanding of the nonperturbative aspects of string 
theory \cite{polch}
from which the idea of so called M-theory, which unifies the 5 known types of 
string theory
has emerged, have continued to have a major impact on string inspired 
phenomenology.
M-theory phenomenology  is now a rapidly growing  field  which, perhaps for the 
first time,
 is allowing us to investigate detailed phenomenological questions in a 
consistent framework 
at large string coupling. There  is  considerable promise in models derived from 
strongly coupled $E_8 \times E_8 $ heterotic strings which in our present 
understanding
is realized through the Horava-Witten (H-W) construction of   $d=11$ 
supergravity compactified
 in a consistent way to $d=10$  on $S^1/Z_2 $  \cite{wh1}, \cite{strw}, 
\cite{wh}. This model 
 provides a natural resolution to the old 
puzzle that was a feature of perturbative heterotic theory, namely the 
consistency problems 
that arise 
if we demand that  reasonable values for $M_{pl} , M_{GUT} $ and $\alpha_{GUT} $ 
should
 emerge in the effective $d=4 $ theory. This resolution hinges on the fact that 
in the 
 M-theory picture the effective string coupling is related to the size of the 
line element
  $S^1/Z_2 $ which is large compared to the $d=11$ Planck length. At the same 
time such a
   picture can naturally 
accommodate gauge coupling constant unification at the experimentally favoured 
value of 
$ 2 \times 10^{16}\; GeV $\cite{strw}. There is a growing body of work dealing with 
phenomenological aspects of  compactified  HW theory. 
Historically this began within the context of the standard embedding  
\cite{msus}  and more recently  
the analysis has been extended to  include nonstandard 
embeddings \cite{nstand} -
\cite{mun} \footnote{Certain aspects
of nonstandard embeddings in M-theory compactifications were already discussed 
in \cite{min}.} 
The correct procedure in obtaining an effective $d=4$ action is to first 
integrate out 
the degrees of freedom on the $6$-dimensional Calabai-Yau manifold and then 
those compactified on  $S^1/Z_2 $  \cite{hpn1}, \cite{lt1}, \cite{mira}, 
\cite{low5} , \cite{elpp1}. 
This new picture provided by the H-W construction  also suggests that the 
mechanism of 
supersymmetry breaking should naturally be explored in $d=5$  rather than the 
usual four 
dimensional setting that is familiar in the weakly coupled case. This follows 
because consistency 
requires that  the scale of 
$S^1/Z_2 $  be hierarchically larger than the typical scale of  the six 
dimensional compact 
manifiold which reduces the theory from $d=11$ to $d=5$ \cite{strw}. The $E_8$ 
and $E_8'$ 
sectors of the four dimensional world then live on the two  boundaries of this 
$d=5$ supergravity
 model. Below the mass scale $m_5 = 1/R_5$ it is adequate to describe the 
low energy physics within the framework of the 4d $N=1$ supergravity. 
In particular, in this framework one can study the dynamical selection of the 
vacuum, and low energy supersymmetry breaking. The interesting question is 
how the nontrivial $5d$ structure of the gauge    
sector, the fact that it consists of two spatially separated, although 
correlated through the choice of the embedding sectors, becomes reflected in
the low energy effective Lagrangian. 
To get an insight into this problem, it is instructive to study 
supersymmetry breaking in simple nonstandard embedding models in the 
Horava-Witten setup. 

Reducing from $d=5$ to $d=4$ 
 around bulk field configurations that satisfy their equations of motion is a 
natural way 
 of encoding the supersymmetry breaking 
dynamics \cite{low5}, \cite{elpp1}, \cite{elpp2}. 
 For example if one considers supersymmetry breaking via  the formation 
of 
gaugino condensates  on a wall  then one approximation is to consider
 these as generating a delta function source proportional to  $\Lambda^3 $ where 
$\Lambda $
  is the scale at which condensation on that wall. It is then possible in 
certain cases, see for instance
   \cite{elpp1}, \cite{elpp2}, to compute the size and nature of  
supersymmetry breaking among observable fields.

Clearly to obtain definite predictions  for the scale of soft supersymmetry 
breaking parameters  
one needs to know allowed values of $\Lambda $. 
In this paper we want to address this question by considering a number of 
explicit examples 
in which the scales $\Lambda $ may be  computed.  To achieve this requires 
knowledge of the 
massless charged spectrum present in the condensing sector as the latter is 
triggered by the 
running gauge coupling  being driven to values larger than unity. 
For M-theory compactification on smooth  Calabi-Yau (CY) manifolds, as in the 
case of weakly 
coupled strings it is difficult to extract explicit data concerning the massless 
charged 
spectrum in most cases. This is particularly the case 
of  CY compactifications 
whose gauge and tangent bundle structure corresponds to nonstandard 
embeddings 
\cite{dist}.

An alternative approach to the problem is to go to the singular limit of  CY 
compactifications 
and 
work with orbifold compactifications.  In the weak coupling limit these spaces 
allow the 
full computational power of string theory to be employed.
The question arises whether the kinds of data one can compute at weak coupling 
such as 
 details of the massless charged spectrum, 
gauge coupling threshold corrections, structure of  the K\"ahler geometry 
underlying the 
 effective low energy theory in $d=4$  can somehow be extrapolated to the strong 
coupling 
 region. In as much as one can at present only  work within the consistent field 
theory limit
  of M-theory compactified on 
$X \times S^1/Z_2 $ of the HW construction the evidence points to many of the 
features 
familiar from 
the weakly coupled theory surviving the extrapolation to strong coupling.  An 
example of 
 such features are  moduli dependent gauge coupling threshold corrections. In 
\cite{stieb} 
  (see \cite{iba}
for earlier work in the same direction)  extrapolation of the latter to large
string coupling by considering the large $T$ behaviour of  modular invariant 
threshold
 corrections \cite{kl} , \cite{ns} 
results in a form of the holomorphic $f$ function that is consistent with the 
one based
 on the supersymmetrization of Witten's threshold formula obtained from M-theory 
\cite{strw}. 
 This was originally carried out for orbifolds with the standard embedding, but  
has recently 
  a more general analysis  has been applied to the case of nonstandard 
embeddings 
  \cite{stieb}. 
  The results agree with the generalization of  Witten's threshold formula to 
include nonstandard 
  embeddings \cite{nstand}.   These extrapolations were 
based on modular invariant perturbative string models  which match onto  HW 
theory
 with nonstandard embeddings but without $5$-branes in the bulk. Such 
extrapolations 
 may also exist 
in the case where such $5$-branes are included \cite{low4}, \cite{low6}, 
\cite{mun}, 
 but an explicit calculation is
 problematic as the 
corresponding weakly coupled strings are not modular invariant \cite{stieb}.

In this paper we consider  nonstandard embeddings in HW theory without $5$-
branes
in the bulk,  where the internal compact space is taken to be an $N=1 $ 
orbifold.
For the purpose of this paper we shall also assume that the 
renormalization group running of the  $d=4$ gauge couplings  on each wall to be 
determined by 
charged massless states living on the wall, the latter  being determined by 
the modular invariant orbifold construction at weak coupling.

\section{Condensates on Opposite Walls}

Following the reference \cite{elpt} we shall summarize the 
main features of the racetrack models with condensates forming in 
two different gauge sectors, with different gauge kinetic functions. 

To begin with let us, however, comment on the connection 
between higher (5d) and four dimensional field theory 
in the 
H-W model. This issue has been studied in detail in \cite{elpt}. 
Here we shall 
recall briefly why do we expect the usual four dimensional physics to play 
the crucial role in supersymmetry breaking and moduli stabilization. 
The general  equation of motion for
a $Z_2$-even field $\varphi$, with the sources ${\cal J}_v(x),\; 
{\cal J}_h(x)$ 
localized on the walls and with the corresponding
boundary conditions on the half-circle \beq \lim_{x^5 \rightarrow
0} \varphi'= \frac{{\cal J}_{v}}{2},\; \lim_{x^5 \rightarrow \pi \rho}
\varphi'=-\frac{{\cal J}_{h}}{2} \label{ssrcb} \eeq  has
the solution of the form $\varphi(x,x^5) = \varphi(x) + \phi(x,x^5)$.
In that
expression $\varphi(x)$ is a truly four dimensional fluctuation, the
zero mode on $S^1 /Z_2$,
 whereas
$\phi$ is the background depending on $x^5$, in the lowest order
approximation given by \cite{llo,elpp1} \beq \phi(x,x^5) = -
\frac{{\cal J}_{v} + {\cal J}_{h}}{2 \pi \rho} ( \frac{(x^5 )^2}{2} -
\frac{\pi \rho^2 }{6} ) + \frac{{\cal J}_{v} }{2} ( x^5 - \frac{
\pi \rho }{2} ). \eeq  This {\em does not} depend
on the particular form of embedding, as explained in \cite{nstand}, but 
it relies on the assumption, that the sources ${\cal J}_v, \; {\cal J}_h$ 
vary significantly only over the scales ${} > \; \pi \rho$, see \cite{llo} 
for details. This condition, on which the form of the explicit solutions 
of the Bianchi identity relies, is nontrivial, as the sources contain not only 
vacuum configurations, but also include all the 4d fluctuations of fields 
which penetrate the walls\footnote{After the averaging over the compact 6d space has been performed.}. However, the assumption about scales is fulfilled 
when one is solving for the vacuum configurations, which might break 
supersymmetry, but should not break the 4d Lorentz invariance, so cannot 
depend on $x^\mu$. At vacuum, the function $\phi(x,x^5)$ is determined 
in terms of the vacuum values of the sources, and perhaps by nontrivial 
physics in the bulk, like sigma-model interactions, but  
the zero mode $\varphi(x)$ , with no dependence on $x^5$, is left completely 
undetermined. This mode
shall eventually be 
regarded as the massless field in 4d, and has to be determined by nontrivial 
4d dynamics born on the walls. Let us note, that even in the schemes where 
bulk dynamics is credited for stabilization of some of the moduli like 
the radius of the fifth dimension, see 
\cite{wise}, a nontrivial potential on the walls is necessary for 
stabilization\footnote{For earlier arguments similar to that of 
\cite{wise} see \cite{elpp1}. The nontrivial warp factor used explicitly in 
\cite{wise} is strictly speaking a higher order correction from the point of view of \cite{elpp1}, since boundary sources are already corrections in the expansion in powers of $\kappa^{2/3}$ (where $\kappa $ is the $d=11$ gravitational coupling).}.

After this introductory comment, we can move on to the discussion of condensates, which are the obvious source of stabilizing potentials and/or hierarchical 
supersymmetry breaking.
To be reasonably general let us start with an arbitrary number of condensates 
on any of the two walls (see also discussion in \cite{dine}). 
With 4d gauge kinetic functions of the 
form $f_{1,2} = S \pm \xi_0 T$ the effective potential in this case 
is 
\beq
V= | -\frac{e^{K/2} W}{M} + \frac{S + \bar{S}}{4 M} ( \sum_i \Lambda^{3}_{i}
) |^2 + \frac{1}{3} | - 3 \frac{e^{K/2} W}{M} + 
\xi_0 \frac{T + \bar{T}}{4 M} ( \sum_i \epsilon_i 
\Lambda^{3}_{i}) |^2 - \frac{3}{M^2} | e^{K/2} W |^2 \label{pmany}
\eeq
where $\epsilon_{i} = \pm 1$ and the rest of the notation is standard
and given in \cite{elpt}.
First, let us look for the flat space supersymmetric points of the potential 
(\ref{pmany}). One reason is purely technical, namely it is much easier 
to find  such candidate points than to look for  broken supersymmetry 
solutions to the full equations of motion. Secondly, as we expect 
the realistic supersymmetry breaking scale to be hierarchically smaller 
than the Planck scale, 
one can expect the relevant points where supersymmetry is only slightly 
broken, to be located near the globally supersymmetric points 
(although the existence of remote relevant susy breaking points 
cannot be excluded in general). 
As we are looking for flat space solutions, we shall assume the vacuum 
expectation value of the total superpotential to be zero, which is consistent in 
the picture with explicit gaugino bilinears, as it does not lock to each other  
the values of various condensates. Then, the scalar potential takes the form 
\beq
V= \frac{(S + \bar{S})^2}{16 M^2}  |\sum_i \Lambda^{3}_{i}
 |^2 + \xi_{0}^{2} \frac{(T + \bar{T})^2}{48 M^2} | \sum_i \epsilon_i 
\Lambda^{3}_{i} |^2 \label{ppmany}
\eeq
and is equivalent to the potential obtained in the globally 
supersymmetric limit in the effective superpotential approach to 
multiple condensates. It is easy to see that existence of supersymmetric 
minima implies that the sum of condensates vanishes separately on each wall
\beq
\sum \Lambda_{i+}^3 = 0 = \sum \Lambda_{i-}^3
\label{eqwalls}
\eeq
at such points in field space. Here $i+, i- $ run over the number of 
condensates present on wall 1 and wall 2 respectively and $\{i \} = \{i+, i- 
\}$.  

In the case of at least two condensates on each wall, equations 
(\ref{eqwalls}) are two independent equations for two complex variables 
$e^{S \pm \xi_0 T}$, so generically they have a solution at finite values 
of $S$ and $T$. The situation is such, that condensates on each wall 
optimize themselves and supersymmetry is unbroken, but the values of 
$S$ and $T$ become fixed. 
This is an interesting possibility, and it would be an interesting exercise 
to check which values of $Re(S)$ and $Re(T)$ can be obtained in such a setup,
but in this paper we want to stay specifically within the class of 
calculable perturbative nonstandard embeddings discussed in \cite{stieb}, 
and there doesn't seem to be enough space for ${} \geq 4$ condensates with 
realistically low condensation scales. 

Hence  we have to look at the vacua with 3 or less condensates. 
These are the interesting cases, given our comments above. 
First, when we want to have 3 condensates  arranged on different walls, then one 
of them must be on one wall, and two on the 
opposite. Then it follows immediately   that to fulfill the unbroken susy 
conditions the single condensate would have to vanish. With three 
nonvanishing condensates on both walls we therefore always have 
broken supersymmetry. The same applies, 
as pointed out in \cite{elpt}, to the case of two condensates on 
different walls. 

When we have several condensates on a single wall, then this 
works similarly to the weakly coupled case: supersymmetry is unbroken, 
but only $S+ \xi_0 T$ or $S - \xi_0 T$ is fixed. 
To have a hope to fix 
both moduli within the simple, and perhaps most appealing, 
version of the racetrack scheme one clearly needs 
to consider condensates on both walls.
    
To conclude, in the most interesting case of two or three condensates 
on different walls, we can exclude the existence of  flat space, 
unbroken supersymmetric ground states. This is not so bad however, 
because the existence of remote minima, disconnected from any globally 
supersymmetric state, cannot be excluded by a general reasoning. 
In fact, if one finds any proper minimum of the effective potential in 
these cases, supersymmetry is guaranteed to be broken there. However, 
to have a vanishing cosmological constant at such a minimum, one has 
to invoke a nonzero expectation value of the superpotential, and it would 
have to be of the order of $\Lambda^{3}_{eff}$ if we are to get the usual 
 hierarchy $F \approx \Lambda^{3}_{eff}/M$. One source of such 
effectively constant 
terms in the superpotential could be condensates of the Chern-Simons forms,
but then the question of the scale and `stiffness' of these condensates 
arises, which we shall not discuss here. It is worth noting  that even if the  
constant in the superpotential is present, then on the basis of the 
dynamics described by (\ref{pmany}) one cannot tell whether it makes 
$F_T$ or $F_S$ vanish. In fact, the most plausible situation in such a case is
that supersymmetry is unbroken, moduli fixed, but the cosmological 
constant is nonzero. 

Having discussed issues of moduli stabilization, there is in addition the 
requirement that the condensation scale on the walls be hierarchically smaller 
than  $M_{Pl} $ in order to have a hope of realistic phenomenology. Of course 
it remains a hope that in the interesting cases mentioned above where 
racetrack 
fixing of  moduli occurs, such a hierarchy of scales appears for a choice of 
embeddings. Whether this is possible or not, we can at least within the 
perturbative orbifold framework of nonstandard 
embeddings \cite{stieb} estimate 
the size of condensates in  some simple cases, even if we do not at the same 
time immediately address the issue of moduli stabilization. This will be the 
subject of the next section. In models where
only a single $T$ modulus exists (whose real part is proportional to the 
size of the 5th dimension) the form of $\Lambda $ is tightly constrained 
if one uses as input the preferred values of $\alpha_{GUT}, M_{GUT} $ and $ 
M_{Pl} $. If one can find reasonable values for the condensate in these models 
it would provide further motivation to tackling the larger problem of 
stabilization. 

In the next Section we discuss in detail potentially realistic examples,
where at the string scale, with trivial vacuum for the scalar fields, 
there is just one asymptotically free nonabelian factor. To repeat, the 
analysis is valid 
in the vicinity of the trivial scalar vacuum. However, in these models there 
are more than one nonabelian factors. In addition, usually the 
perturbative scalar potential has flat directions, along which some of the 
large nonabelian groups can be broken in such a way, that there appear  
additional nonabelian factors which are asymptotically free. It is 
straightforward to generalize in this spirit 
the stringy analysis we give here 
 along the lines of \cite{rosspok}. 
This way one 
can easily generate models with two or three condensates.

\section{Scales of Gaugino Condensation in M-theory on $S^1/Z_2 $:
weak and strong wall cases }
As was discussed in
\cite{nstand}  one can make use of the different routes in compactifying 
M-theory to four dimensions, to obtain a map between "strong" and" weakly"
coupled compactification moduli. This involves compactifying from
$11\rightarrow 10 \rightarrow 4 $  or $11 \rightarrow 5 \rightarrow 4 $
The starting metric in $11$ dimensions is the same, but in the first 
case we can naturally express parameters of the  $d=11$ metric in terms of 
familiar moduli of weakly coupled  $10 $ dimensional heterotic string theory
whilst in the second we have a natural expression in terms of strong units.
If for example we consider case of homogeneous Calabi-Yau compactification, 
then in the weak case we have 
$ g_{\mu \nu}^{(10)}  = e^{- 3 \sigma }  g_{\mu \nu}^{(4)} , 
g_{MN}^{(10)} = e^{\sigma } g_{MN}^{(0)} $, $ \mu = 0..3, M = 4..9 $
with $\sigma $ the breathing mode of the homogeneous CY space.
Along with the dilaton $\phi $ this defines the real parts of weak moduli $S_w
$ and $T_w $ as $Re(S_w) = e^{3\sigma } \phi^{-3/4}  $, 
$Re(T_w) = e^{\sigma } \phi^{3/4} $. The $\phi $ field can be thought of as
$g_{11,11}^{(11)} = \phi^2 $ with a Weyl rescaling $g_{AB}^{(11)} \rightarrow 
    \phi^{-1/4} g_{AB}^{(11)} , A = 0..9$ to obtain canonical gravitational
 action in $d=10 $. In the ``strong" route to four dimensions,  one identifies 
moduli through $ g_{m  n }^{(11)} = e^{-2 \beta } g_{mn}^{(11)}  $, 
$ g_{MN}^{(11)} = e^{\beta } g_{MN}^{(11)} $ and $g_{55}^{(5)}  = e^{2 \gamma }$
Equating $g_{11,11}^{(11)}  = g_{55}^{(5)} $ we find the relation 
$\phi = e^{\gamma-\beta } $. Similarly comparing  $g_{MN}^{(11)} $ and 
$g_{MN}^{(5)} $ we obtain $ e ^{\sigma } = \phi^{1/4} e^{\beta} $
In this way one finds a map between "weak" and "strong" moduli 
which leads one to  define $Re(S_s ) = e^{3 \beta} = Re(S_w) (\phi , \sigma )$
and $Re(T_s ) =   e^\gamma =    Re(T_w) (\phi , \sigma ) $where the subscript  
$s$ refers to strong quantities . 

In  the same way one can consider inhomogeneous CY spaces with 
different scaling moduli. In the orbifold limits these moduli are
associated   with the scaling modes of   the  underlying $6- $dimensional 
 torus      as   $\sigma_i $ $i = 1,2,3$ where $i$ labels the 3 complex planes.
If we include the generalized strong moduli $\beta_i $ then a similar analysis
to above yields the relations $\phi = e^{\gamma - \sum_i \beta_i } $
and $e^{\sigma_i} = \phi^{1/4} e^{\beta_i }$ which leads to the definitions
$Re(S_s) = e^{\sum_i {\beta_i}}  $ , $Re(T_{si}) = e^\gamma (e^{\beta_i - 
\frac{1}{3}
\sum_i {\beta_i}} ) $ in this case. 
Such maps between strong and weak moduli allow one to extrapolate various weak
coupling quantities to the strongly coupled regime. One may wonder if such
extrapolations are consistent. In the case where one includes $5$-branes in the 
bulk
\cite{low4} , \cite{low5}, \cite{mun},  there would appear to be problems in 
this respect as
the corresponding weakly coupled  string theory  is not modular invariant
\cite{stieb}. If we consider the case where such $5$-branes are absent
(which  we assume in the remainder of the paper) the answer seems
more encouraging particularly
regarding  the 
moduli dependent threshold corrections. Computing these quantities in the M-
theory regime
confirms that they can be obtained by the maps considered above as
if we  additionally take the  large $T_i$ limit. The latter ensures that the 
moduli dependent
thresholds are at most linear in the fields, which is within the Horava-Witten 
approximation
to M-theory. Thus in what follows we will assume the perturbative results for 
threshold
corrections in the context of N=1 orbifold compactifications and apply the above
mentioned extrapolations. In what follows we shall drop the $s$ subscript on all 
moduli.

We begin with a discussion of threshold effects in $E_8 \times E_8 $ heterotic
M-theory. 
In the case of smooth Calabi-Yau compactification, the difference in the
unified gauge couplings $\alpha_h^{-1}  $ and $\alpha_v^{-1} $ on the hidden and 
visible walls, 
for general embedding \cite{nstand} can be obtained as a generalization of the 
result due to Witten 
 \cite{strw} is 
\beqa \label{eq:1}
\alpha_{h}^{-1} - \alpha_{v}^{-1} &=& - \frac{s_i}{4 \pi^2 }
 T (n_{F \, i} - \frac{1}{2} n_R ) 
\eeqa
where the single $T = e^{\gamma}$ modulus appears if we consider only the
overall   $(1,1) $ modulus of the CY manifold. The  instanton numbers 
 $n_{F\, i} $ and $n_{R} $ are defined as  
\beq\label{top}
\int_X \omega \, \wedge \,  tr(F^{(i)} \wedge F^{(i)}) = 
-\frac{1}{2} \, \int_X tr F^{(i)}_{ij} F^{(i)\,ij}
 = - 4 \pi^2 \, n_{F\, i}\, \leq \, 0
\eeq
and 
\beq\label{top2}
\int_X \omega \, \wedge \,  tr(R \wedge R) = 
-\frac{1}{2} \, \int_X tr(R_{ij} R^{ij}) = - 4 \pi^2 \, n_{R}\, \leq \, 0 \, .
\eeq
The integrability conditions for the equations of motion give constraint on 
the instanton numbers\footnote{In fact, the configurations we use here 
fulfill the Yang-Mills equations of motion and Einstein equations which 
justifies
 the term instanton.} \cite{gsw}
\beq 
 n_{F\, 1} +  n_{F\, 2} = n_{R} \, . 
\eeq
Let us take standard embedding first. There $n_{F \, v} = n_R$, and
\beqa \label{eq:2}
\alpha_{v}^{-1} - \alpha_{h}^{-1} &=& \frac{1}{32 \pi^3 }
 T n_R  \, .
\eeqa
This particular embedding gives the specific gauge group structure 
$E_{6 (v)} \times E_{8 (h)}$. We stress that since $n_R$ has  
positive sign, there is no way of changing the sign of $\alpha_{v}^{-1} - 
\alpha_{h}^{-1}$ without going to an entirely
 different gauge group structure, which means going to a new, 
different from the standard one, embedding. 
In the $S$ and $T$ notation we have ($\alpha' = 1/2 $) 
\beq \label{eq:3}
\alpha_{v}^{-1} = 4 \pi (S + \epsilon \, T ), \;\quad
 \alpha_{h}^{-1} = 4 \pi ( S - \epsilon \, T ) ,\; \quad
\epsilon = \frac{n_{F \, v} - \frac{1}{2} n_R}{32 \pi^3}
\eeq

where $ S = 1/(g_4^2)  - i \frac{\theta}{8 \pi^2 } $ and $T = r + i
\Sigma $ are defined such that the correct axionic shift invariances of the
low energy $d=4 $ action, in the presence of instantons, is 
$\theta \rightarrow \theta + 2n \pi $ and $\Sigma \rightarrow \Sigma + 16
\pi^2 $.
The equation (\ref{eq:3}) holds for general embeddings. The labels $v$ and $h$
pertain to the standard embedding case where standard model gauge group and 
matter
representations are associated with the  $E_6 $ sector. However it may well be 
that 
for certain embeddings the standard model matter might emerge from either wall, 
thus leading to the possibility that visible matter has strongest gauge coupling 
at $M_{GUT} $ (explicit examples are given later).
We continue to use the labeling  $v$ and $h$ in the general case but the latter 
point should be 
borne in mind.

The above threshold formulae  correspond to following  gauge
coupling functions 
\beq\label{eq:4}
f_h = S - \frac{(n_{F_v} - \frac{1}{2} n_R )}{32 \pi^3} T \qquad , \qquad 
f_v = S + \frac{(n_{F_v} - \frac{1}{2} n_R )} {32 \pi^3}  T  \, .
\eeq

So far these thresholds apply in the case of smooth CY spaces.  We are 
interested
in this paper in the orbifold limit, in which case we need the analogue of
equations (\ref{top}), (\ref{top2}). In \cite{stieb} this problem was addressed within 
the 
context of nonstandard embeddings and the M-theory limit. It was argued that 
topological integrals in eqs(\ref{top}),(\ref{top2}) would only be nonvanishing 
if
there are codimension 2 fixed points appearing. From this it was argued that 
effectively dynamics of $K_3  \times T^2 $ compactification  controls the 
non-zero  $SU_2 $ instanton numbers. The orbifold limit then corresponds to  
taking $K_3 $ orbifolds $T^4 /G $ where  $ G  $ is  an appropriate 
discrete group.  The computation of  allowed instanton numbers  of 
gauge and tangent bundles over  $K_3$ is simplified if the point like 
singularities 
of $T^4/G $ are repaired  by so called ALE spaces \cite{stieb}. In this case the 
$SU_2 $ instanton numbers $n^{(1)}_i , n^{(2)}_i $  are given by 

\beq\label{sutoo}
\int_{C_i} Tr (  F^{(1)} \wedge F^{(1)}  ) =   n^{(1)}_i  \qquad 
\int_{C_i} Tr ( F^{(2)} \wedge F^{(2)} ) = n^{(2)}_i 
\eeq
where $C_i $ are 4-cycles  associated with codimension 2 fixed points 
labelled by $i = 1 ...h_{1,1}$ defined so that 
$\int_{C_i} d^j = \delta_i^j $ and $ \int_{X} {J}_i \wedge d^j = 
\delta^j_i $ where $d^i $ are basis of harmonic
$(2,2) $ forms on the CY space  $X$ and $J_i $ a basis for $(1,1) $ forms.
One can further express $n^{(1)}_i , n^{(2)}_i $ in terms of the orbifold
data such as order of fixed points $\nu_\alpha $ and shifts $\gamma_\alpha^I $
defined by the $N=2 $ embedding \cite{stieb}. 
Because of the $K_3 \times T^2 $ structure, the Bianchi identities
imply that $n^{(1)}_i +n^{(2)}_i  = \chi_{K_3} = 24 $. Thus in going to the 
orbifold limit of CY spaces we are forced to consider a more restrictive set of
instanton numbers than in the smooth case. The corresponding expressions 
for the M-theory threshold corrections, analogous to the equations (\ref{eq:4})
will be \cite{stieb}

\beq\label{eq:orb}
f_h = S + \frac{1}{32 \pi^3}\sum_i \frac{|D_i|}{|D|} ( n^{(2)}_i -12 )\,  T^i 
\qquad ,\qquad 
f_v = S +  \frac{1}{32 \pi^3 }\sum_i \frac{|D_i|}{|D| } ( n^{(1)}_i -12 ) \,  
T^i  \, .
\eeq

 The observation that it is $K_3 \times T^2 $ dynamics that
controls the thresholds corrections via the $SU_2 $ instanton numbers
is completely consistent with the perturbative picture of threshold corrections
which are also associated with the existence of $N=2 $ supersymmetric fixed 
planes \cite{kl}, \cite{dkl} .

Let us contrast the above threshold corrections 
 to the following exact form of the holomorphic couplings in
the large $T^i $ limit of orbifold compactification in $E_8 \times E_8 $
heterotic string \cite{kl}. This is obtained assuming
general case of $h_{1,1} $ different $T$ moduli and general levels $k_v , k_h $ 
and assuming our present notation for sectors labelled by ``$h$" and ``$v$" :

\beq \label{eq:5}
f_h = k_h \tilde{S} + \frac{1}{48 \pi} \sum_i \gamma_h^i \tilde{T}^i \qquad ,
\qquad
f_v =  k_v  \tilde{S} + \frac{1}{48 \pi} \sum_i \gamma_v^i \tilde{T}^i
\eeq

where 
\beq\label{eq:6}
 \gamma_a^i =  \sum _I T_a(R_I) (1 + 2 n_I^i) - C(G_a) \, ,
\eeq 
and $I$ runs over the dimension of the matter field irrep $R_I$ of group
factor  $G_a $. $T_a (R_I) = T{r_{R_I}}({(t^a )}^2) $ for any generator $t^a $
of $G_a $ . $C(G_a) $  is the same quantity but with $R_I$ given by adjoint
representation.
The numbers $n_I^i $ define the so called modular weights of matter fields
transforming in $R_I$ irrep, with respect to $T^i$ duality transformations.
(Note $n_I^i $ is the negative of the modular weights $q_I^i $ defined in 
\cite{kl}.)

As was pointed out in  \cite{kl}  , there is another expression for the
coefficients $\gamma_a^i $ whose validity ensures that the field and string
theoretical threshold calculations agree. This has the form (for general
levels $k_h $ and $k_v $) :
\beq\label{eq:7}
\gamma_a^i =  \frac{b_a^{N=2} (i)}{|D|/|D_i|} +  k_a \delta_{GS}^i
\eeq
where in this formula, $b_a^{N=2} (i) $ is the $N=2$ beta function
associated with massless spectrum originating from the twisted sector
having the invariant plane labelled by $i$. $D$ is  the
orbifold point group and $D_i$ the little group corresponding to the
invariant plane.

Using both these forms of $\gamma_a^i $ we can derive an expression for the
$N=1$ beta function $b_a = - 3 C(G_a) + \sum_I T_a (R_I) $ that will be
useful in the formula for the  scale of gaugino condensation 

\beq\label{eq:8}
\frac{1}{3} b_a =  
\frac{b_a^{N=2} (i)}{|D|/|D_i|} -  \sum _I T_a(R_I) (2/3 + 2 n_I^i)+  k_a 
\delta_{GS}^i \, .
\eeq

The definition of the $T$ moduli  are such  that $T^i = r^i + i \theta^i $ with 
the axionic
shift invariance given by $\theta^i \rightarrow \theta^i  - 4 n^i $ , $n^i
$ integers. Notice that in this basis the $T^i $ dependent thresholds  do
not just differ by opposite signs. However by redefining the $\tilde{S}$
field to a new field $S'$
through $T^i $ dependent shifts one can always bring these into the form
directly comparable with (\ref{eq:orb})  (but for generic levels $k_h , k_v $ 
)
\footnote{In the basis where $S $ is not shifted, there are additional 
holomorphic, universal corrections to $ f_h $ and $f_v$ that yield
the antisymmetric nature of the threshold corrections  \cite{ns}, 
\cite{stieb}.} \beq \label{eq:9}
f_h = k_h S' + \frac{1}{96 \pi} \sum_i (\gamma_h^i - \gamma_v^i) \tilde{T}^i
\qquad ,\qquad
f_v =  k_v  S' - \frac{1}{96 \pi} \sum_i (\gamma_h^i-\gamma_v^i) \tilde{T}^i \, .
\eeq

By comparing the latter formula with the previous M-theory thresholds in
the case of nonstandard embeddings, and taking into account the different
normalizations of $T$ moduli we find that in the orbifold limit, and for a
single $T$ modulus (labelled by some fixed value of $i$ )

\beq \label{eq:10}
n^{(1)}_i - 12   = \frac{|D|}{12| D_i |} ( \gamma_v^i - \gamma_h^i )
\eeq

where in  eqn(\ref{eq:10}) the $v$ and $h$ subscripts generically refer to the
$visible$ or weaker sector and the hidden or stronger sector
respectively. If these sectors are direct products of simple groups then 
there is a corresponding value of $\gamma_a^i $ for each such group factor.

Now in the orbifold limit, expressions for $b_v^{N=2}$ and $b_h^{N=2} $
has been obtained in \cite{stieb} by extrapolation of weak coupling
expressions to the M-theory  domain. These depend  on instanton numbers of
gauge field congfigurations over $K_3$ since 
in the orbifold limit the threshold corrections are essentially based on
$K_3 \times T^2 $ dynamics even though the full theory involves 
compactification on a Calabi-Yau manifold giving $N=1$ supersymmetry in
$d=4$. Defining $ n^{(1)}_i = 12 - n_i $ and $n^{(2)}_i = 12 +n_i $ 
the expressions for the $N=2$ beta functions are \cite{stieb} 
 
\beq \label{eq:11}
b_v^{N=2} = 12 - 6 n_i  \qquad , \qquad    b_h^{N=2} = 12 + 6 n_i
\eeq
(here we assume no Wilson line breaking).

Note that there is no explicit group index appearing in the above
definitions, so that if the unbroken gauge group in the four dimensional
effective theory is indeed a product of simple factors (in either 
the weak or strong sector) then the expressions for the $N=2$ beta
functions must coincide for each group factor. This can be seen in a
number of explicit examples discussed in \cite{stieb} and also is implicit
in the nonstandard embedding orbifold examples in \cite{kl} .
The holomorphic $f$ functions can be written as 

\beqa \label{eq:12}
f_h & = & k_h S + \frac{1}{32 \pi^3}  (  \frac{|D_i|}{|D|} n_i + (k_v -
k_h) \delta_{GS}^i ) \, T^i \cr
&& \cr&&\cr
f_v  & = &  k_v S - \frac{1}{32 \pi^3} ( \frac{|D_i|}{|D|} n_i + (k_v -
k_h) \delta_{GS}^i ) \, T^i
\eeqa

where in these equations we have returned to the definition of $S$ and $T^i$
 moduli with the normalizations given earlier.

Before we consider the specific scale of  gaugino condensation,
 we can write down a general form for the $N=1$ beta functions 
that will appear in the expression for gaugino condensates, using previous
relations,
 again in the  extrapolated orbifold limit:

\beqa \label{eq:13}
  b_v &=& 3 \frac{|D_i|}{|D|} (12 - 6 n_i ) - 
  3\sum _I T_v(R_I) (2/3 + 2 n_I^i)+ 3 k_v \delta_{GS}^i \cr
&&\cr
&&\cr
 b_h &=& 3 \frac{|D_i|}{|D|} (12 + 6 n_i ) - 
  3\sum _I' T_h(R'_{I'}) (2/3 + 2 {n'}_{I'}^i)+ 3 k_h \delta_{GS}^i
\eeqa

where  the apostrophe $'$ distinguishes matter representations  coming from hidden sector.
Again we emphasise that in these formulae, the $i$ labels a single
fixed plane under the action of the orbifold point group.

We can now write the following general expressions for the $v$ and $h$ sector
gaugino condensates $\Lambda_v , \Lambda_h $. In the case of $v$ sector
condensation,  (we set $k_v = k_h = 1 $ for simplicity)
\beq\label{eq:14}
 \Lambda_v  =  M_{GUT} \, ( \frac{\alpha^{-1}_{GUT} }{ \alpha^{-1}_{GUT} + 8 \pi 
\eps T_r}
{)}^{\frac{1}{6}} \, e^{ \frac{2 \pi}{b_v}  ( {\displaystyle \alpha^{-1}_{GUT} + 
8 \pi \eps T_r  } ) } 
\eeq
and in the case of $h$ sector condensation, 
\beq \label{eq:15}
\Lambda_h  =  M_{GUT}\,  ( \frac{\alpha^{-1}_{GUT} }{ \alpha^{-1}_{GUT} - 8 \pi 
\eps T_r}
{)}^{\frac{1}{6}} \, e^{ \frac{2 \pi}{b_h}   ( {\displaystyle \alpha^{-1}_{GUT} 
- 8 \pi \eps T_r } ) }
\eeq
where in both these expressions $\eps =   \frac{1}{32 \pi^3}  (  
\frac{|D_i|}{|D|} n_i )$. 
In the examples based on perturbative nonstandard embeddings in orbifolds
which we discuss below, it turns out that 
condensation occurs on only one wall at a time if the vacuum of the charged 
scalar fields is trivial. As was mentioned in the last
section, while this is not the most interesting case from the point of view of 
complete dynamical fixing of moduli, it will  provide a framework in which to 
estimate the scale of soft susy breaking. This simplification of single wall
condensation  means that the
expression for the observable inverse gauge coupling  $\alpha^{-1}_{GUT}
= 4 \pi (S_r - \eps T_r ) $ when condensation occurs on the weaker wall in
the $v$ sector, whereas  if it occurs on the strong wall 
$\alpha^{-1}_{GUT} = 4 \pi (S_r + \eps T_r ) $.
Of course in either case the triggering of condensation requires the
particularly sector to contain  gauge and massless matter fields that lead
to an asymptotically free theory. This  requires a negative $b_v $
or $b_h $ and we shall discuss these conditions shortly. 

Now we note that the lowest order (in  $\kappa^{2/3} $ ) \footnote{
Where $\kappa $ is the $d=11$ gravitational coupling. } fit to the 
4-dimensional Planck mass $M_{pl}$ and GUT scale $M_{GUT} $ gives
\beqa\label{eq:17}
T_r  & = &\frac{M^2_{pl}  ( \alpha_{GUT} { )}^{\frac{1}{3} } }{ 2^{\frac{17}{3}}
    \pi^3 M^2_{GUT} } \cr 
&&\cr    
&&\cr
M_{GUT}^{-6}   & = & 2 \, \pi {\rm Re}S (4 \pi \kappa^2 {)}^{2/3}   
\eeqa
which shows that the vacuum expectation value of  $T_r $ is not a free parameter.
{}From this constraint it can be seen that in general
the scale of the condensate is given as a function of 
$M_{pl} , M_{GUT} ,
\alpha_{GUT} , n_i , \delta_{GS}^i , k_v , k_h $ as well as the matter
field modular weights $  n_I^i , {n'}_{I'}^i $ and Dynkin labels $ T_v (R_I) ,
T_h (R'_{I'})$. Note that if we take $M_{pl} , M_{GUT} , \alpha_{GUT} $ as more
or less fixed by the MSSM (at this point we put aside the possibility of 
so called strong unification \cite{models} ), then the remaining parameters
depend on the specific details of choice of $N=1, d=4 $ orbifold, along with
choice of embedding, Wilson lines, etc.. 

Before looking at some detailed models, it is worth investigating in the
orbifold limit of Horava-Witten M-theory, if one can
realize the scenario discussed in \cite{nstand} that condensation can occur on
either of the two walls, with corresponding observable sectors on the opposite
wall. Firstly, let us note that in the weakly coupled heterotic
compactifications, there appears somewhat different  constraint on the allowed 
values
of modular weights $n_I^i , {n'}_{I'}^i $  depending on whether one considers
large volume smooth Calabi-Yau or singular orbifold compactifications. In
the former \cite{il} it has been noted that calculations of matter field
contributions to the K\"ahler potential show that in this case,  in known examples, the modular
weights are $ n_I^i =  {n'}_{I'}^i = -1/3 $, which is characteristic of
matter fields arising from the untwisted sectors of orbifolds. These values
imply that in the  expressions for the $ N=1 $ beta functions given above, 
the terms corresponding to matter irreps drop out, and one finds 

\beq \label{eq:18}
b_v =   3 \frac{|D_i|}{|D|} (12 - 6 n_i ) + 3 k_v \delta_{GS}^i \quad ,
\qquad \qquad 
b_h =   3 \frac{|D_i|}{|D|} (12 + 6 n_i ) + 3 k_h \delta_{GS}^i
\eeq
where  we have substituted for orbifold expressions for the N=2 beta
functions in each sector, since as argued in \cite{stieb}, the value of these
quantities is not expected to change in going from CY to the orbifold limit.
 Thus in the cases where $\delta_{GS}^i  = 0 $, we find
that there is quite a strong constraint on the allowed values of the beta
functions. Recall that in our choice of embeddings, $n_i < 0 $ (with the
standard embedding corresponding to  $ n_i = -12 $ ), so that it is clear
that condensation can only occur in the $h$ sector under these conditions. 
If we relax $\delta_{GS}^i = 0 $ then since $k_v , k_h $ are both positive 
integers (by assumption that the corresponding affine algebras are unitary)
the only possibility to realize condensation in $v$ sector is to take 
$\delta_{GS}^i < 0 $. 

Even then,  we have to ensure that even if  $\delta_{GS}^i $ is sufficiently 
negative to turn
$b_v $ negative, it must not at the same time turn $b_h $ negative since it is 
the 
$h$ sector where we identify  with the observable sector , e.g. the MSSM where 
we know  that the corresponding beta functions are positive. Thus  it is clear  
that some kind of
hierarchy between the levels $k_v $ and  $k_h $ is needed if this  mechanism is 
to work, 
the question is how large should this be ? and in addition how negative should 
$\delta_{GS}^i$ be?  Evidently we need to constrain $|n_i | $ so that  the terms  
independent
of $\delta_{GS}^i $ in both beta functions is positive , which requires $|n_i |  
< 2 $ or since the 
$n_i $ are integer and negative,  we are lead t specifically $n_i  = -1 $. We 
then find the 
inequalities $| k_h \delta_{GS}^i | < 6 \frac{|D_i |}{| D |}  $ and 
$| k_v \delta_{GS}^i | > 18 \frac{|D_i |}{| D |}  $ to realize $b_v < 0 ,  b_h > 
0 $ i.e.
$ \frac{6}{k_h} > | \delta_{GS}^i |  > \frac{18}{k_v } $. This requires  $k_v > 
3 k_h $.
Although it may seem that one may have arbitrarily large  values of $k_v , k_h $ 
in fact from the formulae given earlier  for $\delta_{GS}^i $  we see that the 
latter are 
integer valued, which clearly bounds  allowable values of the levels.

The  expression for the $v$ sector condensate appropriate to this case,  which 
is a generalization of that 
given earlier to the case where $k_v , k_h $ are different from unity,  is

\beq \label{eq:19}
\Lambda_v  =  M_{GUT}\,  ( \frac{\alpha^{-1}_{GUT} }{ \frac{k_v}{k_h} \alpha^{-
1}_{GUT} 
+ \frac{1}{4 \pi^2 }T_r  ( 1 + \frac{k_v}{k_h} ) } 
{)}^{\frac{1}{6}} \, e^{  (  \frac{ \pi  (  \frac{k_v}{k_h}  \alpha^{-1}_{GUT} +
 \frac{1}{4 \pi^2}T_r  ( 1 +\frac{k_v}{k_h})  ) } { 27 + \frac{3 k_v 
\delta_{GS}^i}{2} }  ) } 
\eeq

where in the above we have substituted for $\epsilon = \frac{1}{32 \pi^3 } $. 
 By examining the scale  $\Lambda_v $ as a function of 
$\alpha_{GUT} $ and $\delta_{GS} $ for various values of $k_v , k_h $  
satisfying
$k_v > 3 k_h $  one can see that reasonable values may be obtained
but the question remains if one can easily find explicit models with
these values of $k_v , k_h $ and $\delta_{GS} $. 

Having discussed the case relevant to large volume Calabi-Yau compactification 
with
all modular weights $n_I^i = n_{I'}^{i'} =  -1/3 $ we want to consider 
compactifications on $N=1$
orbifolds where more general values of the modular weights occur. This means 
that there is a weaker
connection between the $N=2 $ and $N=1$ beta functions since now matter 
representations
contribute to the right hand side of (\ref{eq:13}) . Therefore we are forced 
to consider
explicit orbifold models, where the full $N=1$  charged, massless spectrum is 
known
in which to evaluate  the size of the gaugino condensate . In this paper we look 
at all the
nonstandard embeddings in $Z_2 \times Z_2 $ and $Z_4$ orbifolds
(here we do not consider Wilson lines) and consider what gauge sectors in each 
model
can support gaugino condensation. Also we shall determine (if condensation does 
occur),
whether it falls into the class of "weaker" wall or "stronger" wall condensation 
in the
sense defined earlier. Such a possibility is a consequence of nonstandard 
embeddings.
First we consider $Z_2 \times Z_2 $ models. There are $5$ modular invariant 
embeddings of the
point group into $E_8 \times E_8 $ including the standard embedding giving the 
unbroken gauge
group $E_6 \times U_1^2 \times E_8' $. Details of these models together with the 
full charged
massless spectrum is given in \cite{kl}. The gauge coupling threshold 
corrections in this case will
depend on all three $T_i ., i =1,..3 $  since each complex plane has a 
nontrivial
 little group among the $Z_2 \times Z_2 $ generators. Moreover since each such 
little group
 defines a $Z_2 $, one also has three U-moduli $U_i $, which the threshold 
corrections 
 also depend on (infact in a way which respects $T_i $ , $U_i $ exchange 
symmetry  \cite{kl})
 As usual the $T$-moduli are associated with deformation of K\"ahler class whilst 
$U_i $ 
 modify the complex structure. In determining the allowed values of the gaugino 
condensate,
 we have to take into account the constraint that  connects $M_{pl} $ with 
$M_{GUT} , 
 \alpha_{GUT}$ and other parameters appearing in the M-theory compactification 
to $d=4 $
 In the case of several $T_i $ moduli, the formula for $M_{pl} $
 is generalized to
 \beq\label{eq:20}
 ( \prod_i Re(T_i)  {)}^{1/3}= 
 \frac{M^2_{pl}  ( \alpha_{GUT} { )}^{\frac{1}{3} } }{ 2^{\frac{17}{3}}\pi^3 
M^2_{GUT} }
 \eeq
 which fixes the "overall" modulus  $( \prod_i Re(T_i)  {)}^{1/3} $ 
 in terms of $M_{GUT} , M_{pl} $ and $\alpha_{GUT} $ . Here $M_{GUT} $ 
 is as given in (\ref{eq:17}) with the definition of $S$ generalized to the 
inhomogeneous 
 CY case as discussed earlier.
  But this still leaves a certain
 freedom in the precise values of  $T_i $ . We shall see in a moment that 
certain values
 of $T_i $ might be favoured over others. Note that this relation does not 
involve 
the $U_i $ so its dynamics are less constrained by realistic values for $M_{pl} 
, M_{gut} $
 etc.. Although the threshold corrections do depend on $U_i $ we shall set $U_i 
= 0 , i = 1, ..3 $
 in what follows and comment on the significance of non-zero $U_i $ at the end.
 Table 1 shows  $Z_2 \times Z_2 $ models with nonstandard embeddings, and 
corresponding gauge
 groups. Note that only two of these have sectors which give negative $N=1$ beta 
function
  when one includes all charged massless matter 
 listed in \cite{kl} and hence allow condensation to occur. Table 2 also lists 
the corresponding value 
  of the instanton numbers $n_i $ corresponding to each $ N=2 $ plane,  and 
which dictate the 
explicit moduli dependence of the  gauge coupling threshold corrections 
(see equation (\ref{eq:12})).

As discussed in \cite{kl} there are basically two distinct $N=2 $ gauge groups 
associated with
the possible $N=2$ planes, namely $E_7 \times SU_2 \times E_8' $ or $ E_7 \times 
SU_2 \times SO_{16}' $
 with $N=2 $ massless matter content described in \cite{kl} . 
The surviving $N=1$ gauge group, obtained after GSO projection will be a 
subgroup of either of these. 
Note in $Z_2 \times Z_2 $ the GS coefficients $\delta_{GS}^i  = 0 $ and we take 
levels $k_h $ and $k_v $ for all non-abelian factors to be unity. Also in this 
case
  $|D|/|D_i| = 2 $
for all planes. Although in principle the condensate $\Lambda $ can depend on 
all three 
real moduli ${\rm Re} T_i $ , for now we simplify matters by considering its 
dependence
in the direction $T_1 = T_2 = a \, T_3 \equiv a \, T $ where $a $ is some 
positive real parameter.
Then effectively $\Lambda$ depends  on two of the three real parameters, $a$ and 
${\rm ReT}$. 
The relation (\ref{eq:20}) will  fix ${\rm Re}T$ so that  $\Lambda$ is 
naturally a function of 
the scale $a$ and $\alpha_{GUT} $ assuming $M_{pl} $ is fixed.
 We will comment on more general case later.

 The expressions for  the $f$ functions and condensate scale $\Lambda $ in the
  condensing models 1 
 and 3 of Table 1  (along  this direction in moduli space)  are:
\begin{flushleft}
{\bf  Model 1}
\end{flushleft}
\beqa\label{eq:21}
 f_{E_6} & = & S - \frac{(2a +1)}{16 \pi^3 } \,T \, , \qquad 
f_{SO_8'}  =  S +\frac{(2a +1)}{16 \pi^3 } \, T  \cr
&&\cr
&&\cr
\Lambda_{E_6}  & = & M_{GUT}   ({ 1 
-( \frac{(2a +1)\,M_{pl}^2 \alpha_{GUT}^{2/3} }{ M^2_{GUT} 32 \pi^4 2^{5/3} 
a^{2/3} }  )  
  {)}^{-1/6} \,
e^{  -\frac{\pi }{9}  [ \alpha^{-1}_{GUT} 
-\frac{(2a+1)}{2 \pi^2 } ( \frac{M_{pl}^2 \alpha_{GUT}^{-1/3} }{ M^2_{GUT} 16 
\pi^3 2^{5/3} a^{2/3} } 
 ) ]} }
\eeqa
\begin{flushleft}
{\bf  Model 3}
\end{flushleft}
\beqa\label{eq:22}
 f_{E_7} & = & S - \frac{(2a +3)}{16 \pi^3 } \,T \, , \qquad 
f_{SU_8'}  =  S +\frac{(2a +3)}{16 \pi^3 } \,T  \cr
&&\cr
&&\cr
\Lambda_{E_7}  & = & M_{GUT} ( { 1 
-( \frac{(2a +3)\, M_{pl}^2 \alpha_{GUT}^{2/3} }{ M^2_{GUT} 32 \pi^4 2^{5/3} 
a^{2/3} }  ) 
   {)}^{1/6} \,
e^{  -\frac{\pi}{21}   [ \alpha^{-1}_{GUT} 
-\frac{(2a+3)}{2 \pi^2 } ( \frac{M_{pl}^2 \alpha_{GUT}^{-1/3} }{ M^2_{GUT} 16 
\pi^3 2^{5/3} a^{2/3} }  )
]} }
\eeqa

In obtaining these expressions  we have solved the constraint (\ref{eq:20})
for $T$. Note that in both cases 
the condensing sector has the strongest coupling of the two walls  at the GUT 
scale,
 which can be seen from 
the relative minus signs in the corresponding threshold corrections. 

In Figures 1-2  we have plotted  $\Lambda $ as a function of the scale $a$ and  
$\alpha_{GUT} $, 
where the range of  values of  $\alpha_{GUT} $ are chosen near the MSSM  
preferred value
of $0.04$. What is particularly interesting about both plots is that they show 
that  the condensate
 has a dependence on $a$  which has a distinct minimum. This suggests that the 
dynamics of the condensate 
 might be such that it would adjust itself to favour $a$ close to its minimum. 
 Strictly speaking one should minimize not $\Lambda $ but rather the effective 
potential  
 $V$ given in the Section 2. 
However one can 
check that the
 position of the minima in $T_i $ space is only shifted slightly, with the value 
of $a$ at the minimum
 remaining unchanged, so Figures 1-2 are qualitatively correct. 
 In model 1, the minimum  is close to $a = 1$ which corresponds 
to the "overall 
modulus" direction 
$T^1 = T^2 = T^3 $ in moduli space. The condensate in Model 2
 shows a minimum close to $a = 3 $.  The value of $\Lambda/ M_{pl}$ 
is seen to be about $10^{-5} $ for  model 1 and $7\times 10^{-4} $  in model 3,  
again 
assuming values of $\alpha_{GUT} $ close to $0.04 $.  The scale of $\Lambda $ in 
the latter  
case might seem to be too high for intermediate scale supersymmetry breaking, 
but infact
in this case $\Lambda  $ is very sensitive to the value of $\alpha_{GUT} $ 
chosen.

For example choosing $\alpha_{GUT} = 0.02 $ gives $\Lambda \approx 5 \times 
10^{-6} \,
 M_{pl} $.
Moving on to consider  more generic points in the $T^i $ moduli space, it turns 
out that 
there is a global minimum of $\Lambda_v $ as a function of  $Re( T^2) , Re(T^3) 
$  (solving the constraint 
(\ref{eq:20}) in terms of $Re(T^1 ) $).  These lie approximately along the 
directions 
$T^1 = T^2 = T^3 $ and $T^1 = T^2 = 3 T^3 $ for models 1 and 3 respectively, 
as can be seen  in Figures 3-4. Hence studying  the condensate along such 
directions is natural.

Let us now move on to discuss the case of nonstandard embeddings in $Z_4 $ 
orbifolds.
Again if we do not consider Wilson lines, there are  $12$ inequivelant modular 
invariant embeddings of the
 $Z_4 $ point group generated by $\Theta = (i, i, -1 ) $ into  
$E_8 \times E_8 $ 
 (e.g. see \cite{kob3}).
As usual the moduli dependent threshold corrections  depend on the $N=2 $ 
invariant planes, 
which in this case corresponds to the $i=3 $ plane which has little group $D 
\equiv Z_2 $. As is argued in
 \cite{kl} this fact implies that the GS coefficients
 $\delta_{GS}^i $ are vanishing in all $Z_4 $ orbifolds
as indeed they are for the $Z_2 \times Z_2 $ models discussed above.
  The only nonvanishing instanton numbers are ${n_3} $ 
 and are determined by 
the $N=2 $ gauge group and spectra
corresponding to this fixed plane. Since the little group is again $Z_2 $ there 
are just two  possible 
$N=2 $ supersymmetric sectors which we label type A and B.
 Type A has gauge group $E_7 \times  SU_2 \times E_8' $ and massless charged 
fields consisting of 
  $2 \times ({\bf 56} ,
 {\bf 2}; {\bf 1} ) \oplus 
16 \times ({\bf 56} , {\bf 1}; {\bf 1} ) \oplus 64 \times ({\bf 1}, {\bf 2} ; 
{\bf 1} ) $.  Type B has 
gauge group $ E_7 \times  SU_2 \times SO_{16}' $ and massless charged fields 
consisting of 
 $2 \times ({\bf 56} , {\bf 2}; {\bf 1} ) \oplus 
2 \times ({\bf 1} , {\bf 1}; {\bf 128} ) \oplus 16 \times ({\bf 1}, {\bf 2} ; 
{\bf 16} ) $. 
In addition there are of course the massless vector multiplets that can, after 
the full projection
that breaks $N=2$ to $N=1$,  give rise to charged massless chiral multiplets.
All this means that in  $N=1$ models based on type A, $N=2 $ sectors, the instanton 
number is $n_3 = -12 $ 
and those based on type B have $n_3 = 4 $. Table 2 lists the allowed  
nonstandard embeddings, 
and in each case one can determine by 
inspection which of the models type A or B  they are associated with. To 
determine the exact 
embedding of the $N=1$ gauge group into  $N=2$ requires comparing the $N=1$ 
massless spectra with the 
$N=2$ spectra listed above \footnote{For brevity we have not listed the various  
$N=1$
charged massless states of the models listed in Tables 1 and 2.}. This  
identification is
 important in deducing whether 
the condensing sector is the more strongly or weakly coupled, a fact which has 
important consequences
 that we discuss  later.
For example in the case of model 1, we have $N=1 $ gauge group $ E_6 \times SU_2 
\times E_7'
\times SU_2' $  which is naturally associated with type A,  $N=2 $ gauge group. 
The threshold
corrections relevant in the M-theory region are therefore (we define $T^{i = 3} 
\, = \, T $)
\beqa\label{eq:23}
f_{E_6} & = & S  + \frac{3}{16 \pi^3 }T  =  f_{SU_2}  \cr
&&\cr 
&&\cr
f_{E_7'} & = & S  - \frac{3}{16 \pi^3 }T , \qquad \qquad f_{SU_2}  =  S  - 
\frac{3}{16 \pi^3 }T
 \eeqa
which shows that $E_6 $ and $SU_2 $ gauge sectors are more weakly coupled and in 
this example 
gauginos of the  $E_7'$ sector condense.

An example of an $N=1$ spectra based on type B, $N=2 $ gauge group is in model 2 
of Table 2.
Here we find 
\beqa\label{thresh}
f_{E_6} & = & S  - \frac{1}{16 \pi^3 }T  = f_{SU_2}   \cr
&&\cr 
&&\cr
f_{SO'_{16}} & = & S  + \frac{1}{16 \pi^3 }T \, . 
 \eeqa
Thus in this case the $SO'_{16} $ is the more weakly coupled (at $M_{GUT} $)
 and  is also the sector where condensation occurs. In a similar manner one can
 derive the threshold 
corrections for the models listed in Table 2 and determine the weaker/stronger 
coupled sector. 

In order to compute the $N=1$ beta functions with a view to discovering which 
models can support gaugino
 condensation,
 it is necessary to have knowledge of the complete charged massless spectrum.
In  \cite{kob2}  a classification of the  $ {\cal T}_{k} $
 twisted sectors of $Z_{2k} $,   $N=1 $ orbifolds was presented.
  In the present case of $ Z_4 $ \cite{kob3} list the massless states from the 
${\cal T}_1 $ and ${\cal T}_2 $ twisted sectors. To complete the analysis we
 have just to consider if any additional massless states can arise
 from the twisted sector ${\cal T}_3 $ ($ k = 0..3 $, with ${\cal T}_0 $ 
 corresponding to the untwisted sector in the $Z_4 $ case).
 The equations defining massless  right and left-moving states are
\beqa\label{eq:24}
& \frac{1}{2} & \sum_{t = 1}^4 ( p^t + k\,  v^t  )^2 + N_R^{(k)} + c_k - 
\frac{1}{2} = 0 \cr  
&&\cr
&&\cr
& \frac{1}{2} & \sum_{I= 1}^{16} ( P^I + k\,  V^I  )^2 + N_L^{(k)}  + N_{E_8 
\times E_8'}+ c_k  -1 = 0
\eeqa
where $ N_{R, L}^{(k)} $ and $ N_{E_8 \times E_8'}$ are oscillator numbers in 
the right and left moving degrees.
 
The point group element $\Theta $ acting on the $SO(8) $ bosonized fermion 
lattice is 
represented by the shift vector  given by $  v = \frac{1}{4} ( 
1, 1, -2 , 0) $. The normal ordering constants $c_k $ are $ c_1 = c_3 = 5/16 $ 
and $c_2 = 1/4 $. 
Now since $min( (p+ 3\, v )^2) = 22/16 $  (where $p$  is an $SO(8) $ lattice 
vector) it is clear that there 
are no additional massless modes coming 
from the ${\cal T}_3 $ twisted sector. (In the ${\cal T}_2 $ sector for example 
there are massless states
since now $min((p + 2 \, v ) ^2 ) = 1/4 $.) Thus the end result is that we can 
use the 
states given in \cite{kob3} to determine the $N=1 $ beta functions in $Z_4 $ 
models and those 
which have negative values, and  which are therefore relevant in gaugino 
condensation are listed in Table 2. 
From our above discussion of stronger and weaker coupled sectors at $M_{GUT} $, 
it turns out that all
of the models that support a condensing sector, this sector is the more strongly 
coupled except in two cases: models 2, 4 and 9 where this occurs in the  weaker 
sector.   
Since in the $Z_4 $ case there is only one $T$  modulus contributing to 
thresholds,  the size of the 
condensate (either $ \Lambda_v $ or $\Lambda_h $ as appropriate) is only a 
function of 
$\alpha_{GUT} $  (for fixed  $M_{pl}$ and $M_{GUT} $).

Before closing it is worth while considering what the largest allowable value 
for the radius of the 
$5$th dimension can be in the models considered in this paper, consistent with 
the various constraints
on low energy couplings etc that we have seen above. There has been much recent 
interest in 
models that  macroscopically large  (with respect to e.g. the inverse GUT scale $2 \times 10^{16} $ 
GeV)
extra spatial dimensions. 

One of the basic constraints related 
to the maximum 
physical size of the $5$th dimension,  $\pi \rho  \, = \,  \pi {\rm Re} (T) 
\rho_0  $ 
\footnote{ In \cite{nstand} we argued that a natural choice for the 
reference scale 
$\rho_0  $ is through $\alpha' = \frac{1}{(4 \pi )^{2/3} \pi^2 } \frac{\kappa^{2/3}}{\rho_0}.$} and the 
choice of whether  the observable sector
lies on the wall that is more  weakly/strongly  coupled. In the former case we 
have the relation 
$\alpha_{GUT} = {\rm Re} S + \epsilon {\rm Re}T $ (here for simplicity we 
consider a single modulus $T$) 
 from which the gauge coupling on the hidden wall  can be expressed as  
$\alpha^{-1}_{h} =  \alpha^{-1}_{GUT}
  -2 \, \epsilon {\rm Re}T $. Clearly this choice reveals the critical 
upper bound on ${\rm Re} T $ and the scale of the $5$th dimension $\pi \rho $ 
(for a fixed value 
of  $M_{GUT} $) if we demand
that the hidden sector gauge coupling remain finite \cite{strw}. Although the 
exact value of this
critical radius depends  on the size  of the threshold coefficients $\epsilon $ 
in the models we have
 considered in this paper there will be rough agreement of $\pi \rho =     $ for 
the choice $M_{GUT} = 2 \times 10^{16} GeV $ . The other case one might have 
(which is realized in 
models 1, 4 and 9 of the $Z_4 $ orbifolds listed in Table 2) is that the 
stronger coupled sector contains the 
observable sector in which case we have the GUT scale relation $ \alpha^{-
1}_{GUT} = {\rm Re}S
- \epsilon {\rm Re} T $ and $\alpha^{-1}_h = \alpha^{-1}_{GUT}
+2 \, \epsilon {\rm Re} T $. Thus there is no critical radius in this picture, 
if we imagine 
$\alpha_{GUT} $ as fixed. In both cases however,  ${\rm Re} T $ is determined by 
the requirement
of obtaining the correct value of $M_{pl} $ once $\alpha_{GUT} $ and $M_{GUT} $ 
are fixed. 
But one can imagine (as in \cite{ben}) lowering the value of $M_{GUT} $ from $2 
\times 10^{16} GeV $, 
which would mean obtaining something other than the minimal supersymmetric 
extension of the standard model
in $d=4 $. By lowering $M_{GUT} $ (or in other words  increasing $\kappa $) it 
would seem we  can make 
$\pi \rho  $ larger by not only the increase in $\kappa $ but also of ${\rm Re 
}T $ through the relation 
 (\ref{eq:17}). However if we want to include the mechanism of hidden wall 
gaugino condensation to break 
supersymmetry as we have discussed in this paper, then this provided additional 
bounds on the maximum
values of $\pi \rho $.  This follows from the fact that if the wall on which 
condensation occurs is
the more weakly coupled, then lowering $M_{GUT} $ forces the hidden sector gauge 
coupling at $M_{GUT} $ to 
smaller values and consequently it requires a longer renormalization group 
running (for a given value of 
$ b_h $) in order to trigger condensation. This fact coupled with the lowering 
of  $M_{GUT} $
both point to a lowering of the scale $\Lambda $ at which the gaugino's 
condense. We have seen in this paper
that for the models considered we already obtain values of $\Lambda $ around the 
optimal value of ${10}^{12}-10^{13}
$ GeV for gravity mediated soft supersymmetry breaking with $M_{GUT} =  2 \times 
10^{16} $ GeV.
 A natural lower bound on $M_{GUT} $ would then be that which gives $\Lambda  
\approx 10^{5} $  GeV
 which would correspond to gauge mediated soft supersymmetry breaking. 
{}From our discussions above this 
 provides an upper bound on  $\pi \rho $.
 
  Let us state the results. In the $Z_2 \times Z_2 $ case, and model 1,  taking 
the direction ${\rm Re}T^1 =  {\rm Re}T^2 = a \, {\rm Re} T^3 = a \, {\rm Re}T $ 
(with the dynamically favoured value $a = 1 $ as discussed earlier) we find  the 
critical value ${\rm Re}T_c  \approx 100 $ for $\alpha_{GUT} 
 = 0.04 $ so that the physical critical radius $\rho_c =  16 (2 
\alpha_{GUT})^{1/6} M^{-1}_{GUT} $.
  The actual value of ${\rm Re} T $ obtained from the constraint
 (\ref{eq:20}), with above value of $\alpha_{GUT} $ and $M_{GUT} = 2 \times 
10^{16} \, GeV$ is ${\rm Re}T \approx 78 $ which is  close to the critical value. 
Because of this
 proximity and the fact that ${\rm Re}T$ is very sensitive to $M_{GUT}$ and 
 $\Lambda$ decreasing exponentially fast with $T$ infact we cannot reduce 
$M_{GUT} $
 by much more than a factor of $2$ or so. This implies the upper bound on the 
 physical length $ \rho $ is only a few times that of the critical radius 
  $\rho_c $ given above.  In  model 3  ${\rm Re}T_c  \approx 50 $ for the MSSM
  values of $\alpha_{GUT} , M_{GUT} $. However we have see that at these values 
${\rm Re} T \approx 78 $ and we are in a forbidden region. Infact to remedy this 
one can 
  either decrease $\alpha_{GUT} $slightly or increase $M_{GUT} $  to avoid this 
problem.
   For example in the plot of $\Lambda $ in this case (Fig. 2) we have taken 
$\alpha_{GUT} > 0.02 $.
    
 In the $Z_4 $ case there are two possibilities: either strong or weak wall 
condensation may occur. In the former, the analysis follows as above but now  
critical values of $T$ are ${\rm Re } = 160 $ or $ {\rm Re } = 480 $
depending on whether the particular $N=1$ gauge group originates from either the
type A or B  $N=$ gauge group discussed earlier. These allow for slightly 
smaller reductions in $M_{GUT} $ before ${\rm Re}T$ reaches its critical value 
than in the $Z_2 \times Z_2 $ case being typically $1/5 $ and $1/10$ times the 
MSSM value of $2 \times 10^{16} $ GeV. 
 Potentially the more interesting case is weak wall condensation where the 
critical radius is absent, and it would seem large values of $\rho $ are 
possible.
However the bounds that $\Lambda \geq {10}^5 $ GeV together with the observation 
that
$\Lambda$ decreases exponentially with $T$  which is itself varies as 
$(M_{Pl}/M_{GUT})^2 $ mean that $M_{GUT} $ cannot be lowered significantly.
In models 1 and 9 for example, we typically can only lower by a factor of 
$1/2$ or $1/3$ with respect to the MSSM value, and hence the maximum values of 
the radius $\rho $ are again only a few times the critical values. 

It is clear that since the magnitude of the $N=1$ beta function in the 
condensing sector can never be too large (it is $-90$ for unbroken $E_8 $)
this restriction on the lowering of $M_{GUT} $ and the consequent increase
of $\rho $ is pretty strong and fairly generic. One possible means around this 
would be if one could find threshold corrections  involving several  $T_i$
moduli   
 where the instanton numbers $n_i $ have opposite signs for at least two values
 of the index $i$. Then one can imagine that the above lower bound on 
 $M_{GUT} $could be weakened  if  partial cancellation occurs amongst the moduli
 in the threshold corrections, as we lower $M_{GUT} $ and  hence increase 
 ${\rm Re}T_i $. It would be interesting to find orbifold models where this 
happens and to estimate the maximum allowed value of $\rho$ which might be 
orders of 
 magnitude larger than $\rho_c$. 

As noted at the end of Section 2, it is straightforward to construct models
with two or three condensates using the basis models from Tables 1 and 2. To achieve this,
one needs to identify suitable F-- and D--flat directions in the effective 
field theoretical Lagrangian. Then, typically,  nonvanishing vevs of scalar fields along these flat directions can be chosen in such a way, that there 
remain additional nonabelian sectors with negative beta function 
coefficients, thus providing additional condensates. This can happen for instance to one of the $SO_{8}'$ factors in the model 1 from Table 1, or to the 
$E_6$ factor in the 
model 1 of Table 2. Detailed study of these models requires identification 
of the Yukawa couplings, i.e. of the perturbative superpotential, for all the massless states. This is doable, however complete analysis of this type lies 
beyond the framework of the present paper. Preliminary considerations 
support clearly the scenario outlined in  Section 2.     

 \begin{table}
 \begin{center}
 \begin{tabular}{|c|c|c|c|c|}
 \hline
 \(\strut {{\rm Model}  }\) & \( {\rm Gauge} \,{\rm Group} \quad \)
&\( {\rm Condensing}\, {\rm Sector }\, \tilde{G} \) &
 \(  b_{\tilde{G}} \)&\(  (n_1, n_2, n_3) \)
 \\ \hline
 \(\strut  1 \) & \(  E_6 \times  U_1^2 \times SO_8' \times SO_8' \)  & \( E_6 
\)   & \( -18 \)
 & ( 4, 4, 4)    
\\ 
\(\strut  2 \) & \(  E_6 \times  U_1^2 \times SO_{16}' \)  &  \( {\rm none} \)  
& \(  - \)
 & ( 4, 4, -12)    
\\
\(\strut   3  \) & \(  E_7 \times SU_2 \times SU_8' \times U_1' \)   &\(  E_7 \)  
& \( -42
\) &  (4, 4, 12)
\\
\(\strut  4 \) & \(  SO_{12} \times SU_2 \times \tilde{SU}_2 \times SU_8' \times 
U_1' \) 
 & \(  {\rm none} \)   & \( - \)
 & ( 4, 4, -4)
\\    
\hline
 \end{tabular}
 \end{center}
 \caption{$N=1,  \, Z_2 \times Z_2 $ orbifolds with nonstandard embeddings }
 \label{tab1}
 \end{table}

\begin{table}
 \begin{center}
 \begin{tabular}{|c|c|c|c|c|c|}
 \hline
 \(\strut {{\rm Model}  }\) & \( {\rm Gauge} \,{\rm Group} \quad \)&\( {\rm 
Condensing}\, {\rm Sector }\,
  \tilde{G} \) &
 \(  b_{\tilde{G}} \)&\( \Lambda ({\rm in}\, {\rm units}\, {\rm of}\,  M_{Pl} ) 
\) & \( n_3 \)
 \\ \hline
 \(\strut  1 \) & \(  E_6 \times SU_2 \times E_7' \times SU_2' \)  & \( E_7' \)   
& \( -40 \)
 &$8 \times 10^{-5} $ &-12   
\\ 
 \(\strut   2  \) & \(  E_6 \times SU_2 \times SO_{16}' \)   &\(  SO_{16}' \)  & 
\( -10
\) & $ 2 \times 10^{-5} $&4
\\
\(\strut  3 \) &\(  SO_{14} \times E_7'  \) &\( E_7' \)   &\( -42
 \) &$ 8 \times 10^{-5} $& 4
\\
\(\strut  4 \)& \( SO_{14}   \times  SO_{12}' \times SU_2'   \)& \(  SO_{14}  \)  
& \( -2
 \) &  $ \leq 10^{-10} $ & 4
\\
\(\strut  5 \) &\(  SO_{10} \times SU_4 \times E_7'  \)   &\(  E_7' \)   &\(  -
42
  \) & $7 \times 10^{-5} $&4
\\
\(\strut  6 \) &\(  SO_{10} \times SU_4 \times SO_{12}' \times SU_2' \) & 
\(SO_{12}' \)   & \( -14
 \)&  $ 10^{-7} $&4
\\
\(\strut  7 \) & \(  SU_8 \times SU_2 \times E_8' \)  & \(E_8' \)   & \( -90
 \)& $3 \times {10}^{-3}$ &-12
\\
\(\strut  8 \)&\(   SU_8 \times SU_2 \times E_7' \times SU_2' \)  &\(  E_7' \)   
&\( -42
 \) &$  6 \times 10^{-5} $ &4
\\
\(\strut  9 \) & \( SU_8 \times SU_2 \times SO_{16}' \)   &\( SO_{16}'  \)  & \( 
-26
  \)& $ 1.5 \times 10^{-5} $&4
\\
\(\strut  10 \) & \(  SU_8 \times  E_6' \times SU_2'   \)& \( E_6'   \) & \( -18
 \) &  $ 10^{-6} $&4
\\ 
 \hline
 \end{tabular}
 \end{center}
 \caption{$N=1,   Z_4 $ orbifolds with nonstandard embeddings ($U_1$ factors
suppressed)}
 \label{tab2}
 \end{table}
%
 \section{Soft Terms}
 In this section we want to consider typical expressions for soft supersymmetry
 breaking operators that are induced due to the presence of a gaugino
 condensates on either of the walls, or on both of them. This mechanism is 
naturally described within a 
$5d$
 framework where transmission of supersymmetry breaking between walls occurs
 through
 bulk field interactions with the walls. Although the framework 
of $5d$ supergravity is well
 established, there are  subtleties that arise in the H-W formalism when 
reducing to $d=4$ in the presence of condensates, which were discussed in 
\cite{elpt}. 
 There, it was argued that a method of reduction from $d=5$ to $d=4$ that 
captures
 the dynamics of the H-W picture, is to integrate out (via their equations of 
motion) the bulk fields that 
couple to condensates on (in general) both walls. In this way one obtains 
 the following set of soft operators in the case that condensates 
 $\Lambda_i , \, i =1,2=(+,-)$ appear on opposite walls \cite{elpt} 
 
 \begin{itemize}
 \item {\bf Gravitino Mass} \\
 \beq m^{2}_{3/2} = \frac{(S +
\bar{S})^{2/3}}{2^{2/3} 12 M^4} (\Lambda_{1}^{3} + \Lambda_{2}^{3}
)^2 + \frac{(T + \bar{T})^2}{ 432 M^4} \epsilon^{2} (
\Lambda^{3}_{1} - \Lambda^{3}_{2} )^2 
\eeq 
 where $M$ is the reduced $d=4$ Planck mass
 \item{{\bf Tan}( $\theta  $ )} \\
 \beq \tan ( \theta ) = \sqrt{3} \epsilon^{-1}
\frac{S + \bar{S}}{T + \bar{T} } \frac{\Lambda_{1}^3 +
\Lambda_{2}^3 }{\Lambda_{1}^3 - \Lambda_{2}^3 } 
\eeq 
\item{\bf Gaugino Masses} \\
  Here there is in general, a splitting between the gaugino masses 
 arising on different walls, which we denote by $M^{\pm}_{1/2} $
 \beq
 M^{\pm}_{1/2} = \frac{\sqrt{3} m_{3/2} }{
(S + \bar{S} ) \pm \epsilon ( T + \bar{T} ) } \left ( \sin (\theta)
(S + \bar{S}) \pm \epsilon \cos (\theta ) \frac{(T + \bar{T}
)}{\sqrt{3}} \right )
 \eeq

\item{\bf Trilinear Scalar Interactions} \\
 Assuming vanishing CP-violating phases then generic $A^{\pm}$-term takes the 
form 
 \beqa \label{tri}
  A^{\pm} = &\sqrt{3} & m_{3/2} \left (
\sin (\theta) \left ( -1 \pm \epsilon \frac{3 ( T + \bar{T} )}{3 ( S
+ \bar{S} ) \pm \epsilon ( T + \bar{T})} \right ) \right . \cr
&&\cr 
+ & \sqrt{3} &\cos( \theta ) \left . \left ( -1 +\frac{3 ( T + \bar{T} )}{3 ( S + \bar{S} ) \pm \epsilon ( T + \bar{T}) } \right )  \right ) 
\eeqa
 
 \item{\bf Chiral Scalar Masses} \\
 The chiral scalar masses $ m^{2 \pm}_{i \bar{j}} $ are found to be
 \beqa
& m^{2 \pm}_{i \bar{j}} = K_{i \bar{j}} \left ( m^{2}_{3/2} - 
\frac{3 m^{2}_{3/2} }{3 ( S + \bar{S} ) \pm \epsilon (T + \bar{T}) }
 \left ( \pm \epsilon  ( T + \bar{T}) (2 -
\frac{\pm \epsilon ( T + \bar{T} ) }{3 ( S + \bar{S} ) \pm  \epsilon 
( T + \bar{T} ) } ) \sin^2 \theta  \right . \right . & \nonumber \\
& \left . \left . + (S + \bar{S}) ( 2 - \frac{ 3 ( S + \bar{S} ) }{3 ( S + 
\bar{S} ) 
\pm  \epsilon ( T + \bar{T} ) } ) \cos^2 \theta  -
 \frac{ \pm \epsilon 2 \sqrt{3} (T + \bar{T}) 
(S + \bar{S})}{3 ( S + \bar{S} ) \pm  \epsilon 
( T + \bar{T} ) } \sin \theta \cos \theta \right ) \right ).&
\eeqa
\end{itemize}
If we now consider the single condensate case (e.g.  $ \Lambda_2 = 0$ )
then it is clear from the above expressions that $\Lambda_1 $ sets the scale 
for the soft terms, as well as the appearance of   reduced Planck mass $M$
in  the expression for gravitino masses. It is interesting to see the dependence 
of the various soft terms on
$\epsilon $ ( which  corresponds to a choice of nonstandard embedding), in the 
case where we take standard model values for  $\alpha_{GUT} $ and $ M_{GUT} $,
and eliminate $ReS , ReT $  using the earlier definitions and the fit to
$M_{pl} $ in $d=4 $ . Although  applying this procedure gives soft terms 
that depend on $\epsilon $, plotting these in a model independent fashion
is problematic, since we know that the beta function coefficients in
 $ \Lambda_1 $  change as $\epsilon  $ varies.  A better solution is to  
normalize the plots  by dividing the soft term by appropriate powers of 
$\Lambda_1 $ and $M$  (effectively yielding dimensionless expressions).
 This is what we have done in Figures 5-7 
 \footnote{Note, that we assume the convention that $\theta$
varies between $- \pi /2$ and $ + \pi /2$.} 
 (in the case of chiral scalar masses 
we plot the diagonal  combination  $K^{-1ij} m^{2 \pm}_{jk}  M^4/\Lambda_1^3 $).
 To get absolute  normalization of soft terms in a given model  requires the 
data from the plots  together with the value of  $\Lambda_1 $ in the model.  

The striking feature of the pattern of the soft breaking operators is 
the strongly marked asymmetry between soft terms for fields stemming from 
different walls, both for small and large values of
the $\tan (\theta)$. It is straightforward to imagine the Standard Model 
group to be a mixture of the gauge factors originally\footnote{i.e. before 
turning on the vacuum bundle.} contained in both $E_8$ and $E_{8}'$ 
living on different walls (at least the $U(1)$ factor(s) can easily be such 
mixtures), see \cite{nstand}. In such a case this asymmetry would be a clear 
low energy signature of the `double-wall' structure at high energies.
\section{Conclusions}

In this paper we have focussed on the viability of the mechanism of gaugino 
condensation to generate  supersymmetry breaking
at acceptable energy scales in strongly coupled 
$E_8 \times E_8 $ heterotic string theory.
 In this context we have considered  compactifying the boundary
 theories to $d=4$ on orbifolds, in order to have explicit strongly coupled
string models as the 
 framework to consider the 4d dynamics of the hidden sector condensation.

We have concluded that if two or three condensates are switched 
on {\em different} walls, then supersymmetry is always broken. 
If a larger number of condensates are switched on, then one obtains 
solutions with unbroken supersymmetry, but, possibly, stabilizing 4d
moduli. From the five dimensional point of view, the racetracks on the 
walls are able to fix boundary expectation values of the 5d moduli,
thus giving rise to the scenario described recently in \cite{wise}. 

In all the explicit examples we have studied the scale of 
condensation $\Lambda $ is 
smaller than the scales associated with 
 either the $5$th or orbifold compact dimensions and so justifies
 the treatment of the dynamics of the condensate in the 4d framework. We have 
seen that phenomenologically reasonable 
 values of $\Lambda \approx {10}^{-5} M_{pl} $ are obtainable 
 in a number of models with the MSSM value of 
 $M_{GUT} $ and $\alpha_{GUT} $. In the case of  
 threshold corrections which depend only on a single 
 $T$ modulus, this result is particularly interesting
 as such a modulus is determined in terms of 
 $M_{pl}, M_{GUT}$ and  $\alpha_{GUT} $ by the requirement that 
the  correct value of $M_{pl}$ emerge in $d=4$.
Thus it seems that the  M-theory thresholds are such that 
they appear to favour MSSM values of $M_{GUT}$ in as much as if 
we try to lower significantly the value of $M_{GUT} $ from this, 
then $\Lambda$ is lowered exponentially quickly, due to the 
sensitivity of ${\rm Re}T$ to $M_{GUT} $. The same is 
roughly the case when several moduli are present as we have seen
in the $Z_2 \times Z_2 $ orbifold examples. It would seem that to accommodate 
smaller values of $M_{GUT} $ within models that 
also generate hierarchical hidden wall condensation requires 
some partial cancellation of moduli field expectation values 
in the threshold corrections on that wall, as we discussed above.
    
Even though as we have stated, the magnitude of $\Lambda $ justifies a 4d 
renormalizable  field theory picture, the actual supersymmetry breaking in the 
observable sector appears through soft terms generated by certain 5d bulk 
fields that 
couple to the condensates. 
What we 
have seen in some of the models  studied here is that  supersymmetry breaking 
source terms that arise from gaugino condensation have 
approximately the right
hierarchical mass scale to generate TeV scale supersymmetry breaking
in the visible sector. 
To visualize some of the nontrivial features of this supersymmetry breaking 
mechanism in the presence of two gauge sectors with different gauge kinetic 
functions, we have 
plotted the soft susy breaking operators for fields living on different walls
as a function of the parameter $\epsilon$ which controls the nonuniversal part of the gauge kinetic function. Different values of the  $\epsilon$ 
correspond to different embeddings. 
There is a noteworthy asymmetry between soft terms on different walls,
which might offer a low energy signature of the `double wall' structure of the fundamental theory. 

\vspace{0.5cm}

\noindent{\bf Acknowledgments}: Authors thank Stephan Stieberger for
very helpful discussions.\\
Authors thank also Stefan Pokorski, Hans Peter Nilles and Jan Louis for useful
comments.\\ 
S.T. thanks Physikalishes Institut of the Universit\"at Bonn for 
hospitality during his visit.\\
This work was
partially supported by the European Commission programs
ERBFMRX--CT96--0045 and CT96--0090. Z.L. acknowledges the
support by the Polish Committee for Scientific Research Grant 2
P03B 052 16 (99-2000) and by M. Curie-Sklodowska Foundation,
and S.T. the Royal Society.

\newpage
 \begin{figure}[b]
 \begin{center}
  \epsfig{file=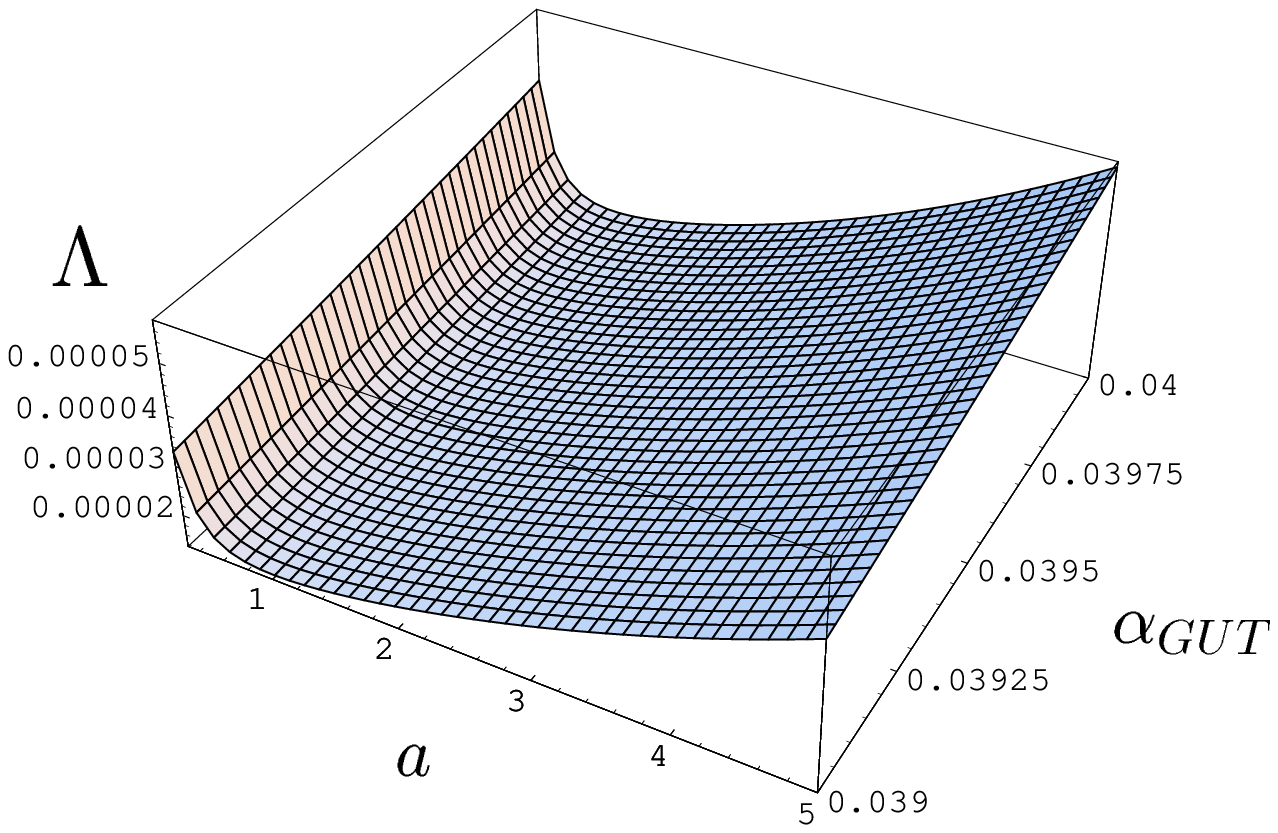, height=8cm} 
  \caption{{\rm  Model 1 } :  $\Lambda $  as a function of $a$  and  $\alpha_{GUT} $. }
\end{center}
  \end{figure}   
  \vspace{2cm}
 \begin{figure}[b]
  \begin{center}
\epsfig{file=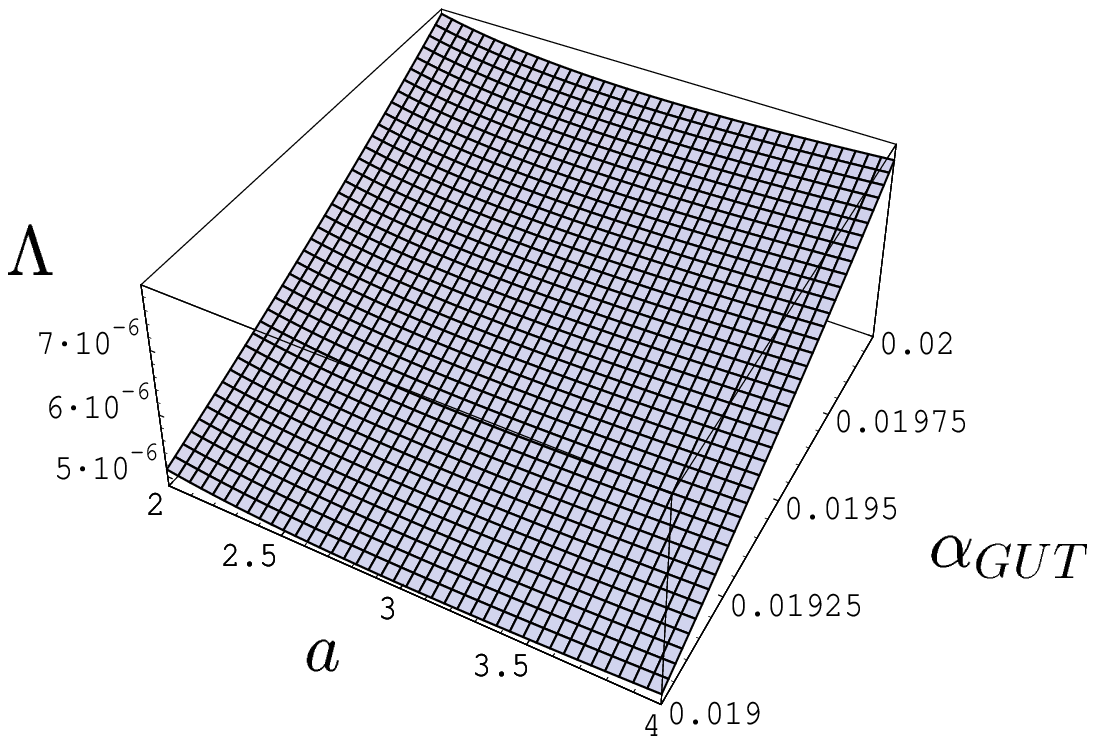, height=10cm}
   \caption{ {\rm Model 3}: \, $\Lambda $ as a function of
 $a$ and $\alpha_{GUT} $.  }
   \end{center}  
   \end{figure}
 \begin{figure}[b]
 \begin{center}
\epsfig{file=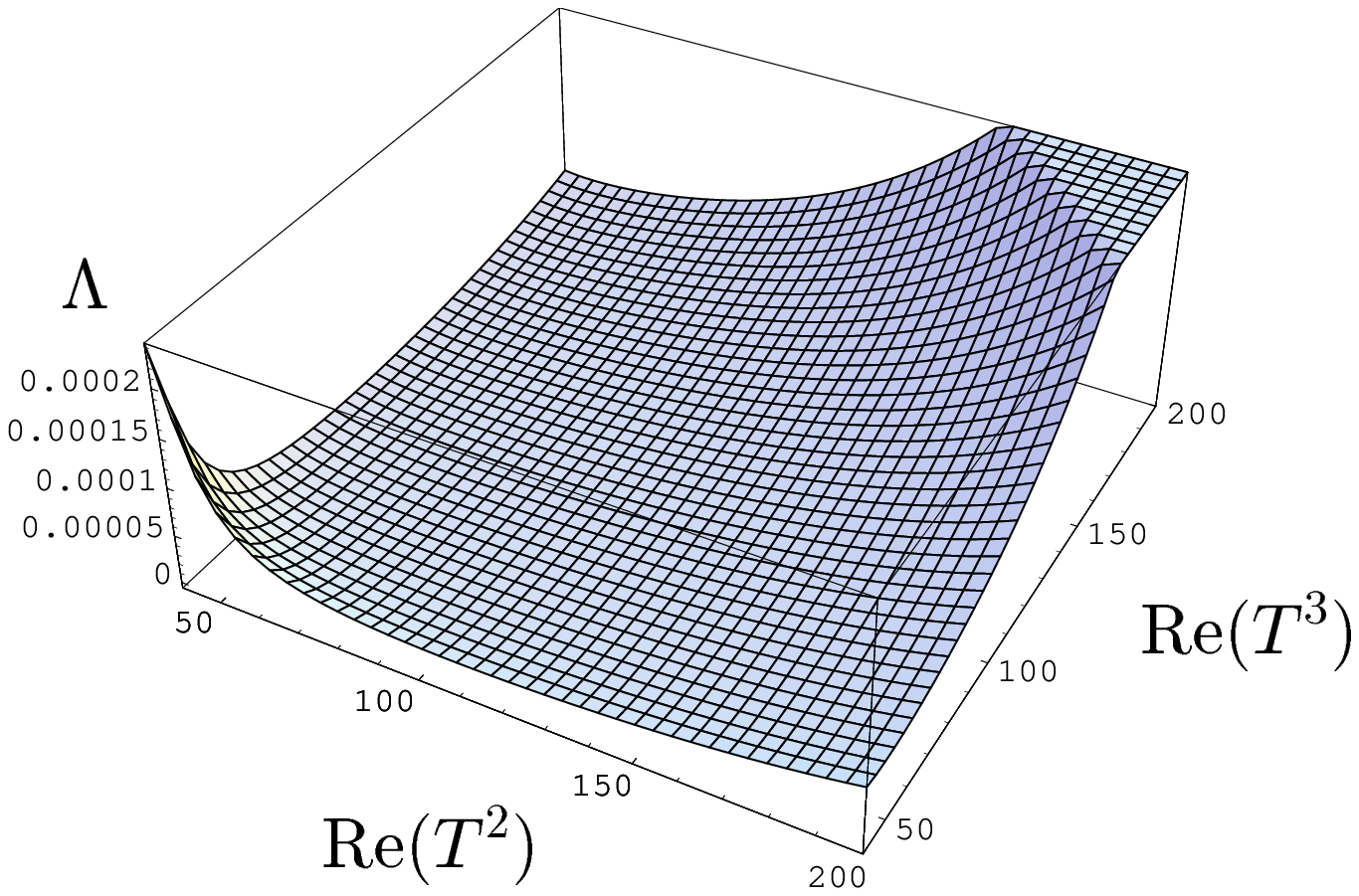, height =7.5cm}
\caption{ {\rm Model 1}: \, Global minimum of $\Lambda $ as a function of 
$ Re(T^2)$ and $Re(T^3)$. }
       \end{center}
        \end{figure}
\begin{figure}[b]
\begin{center}
\epsfig{file=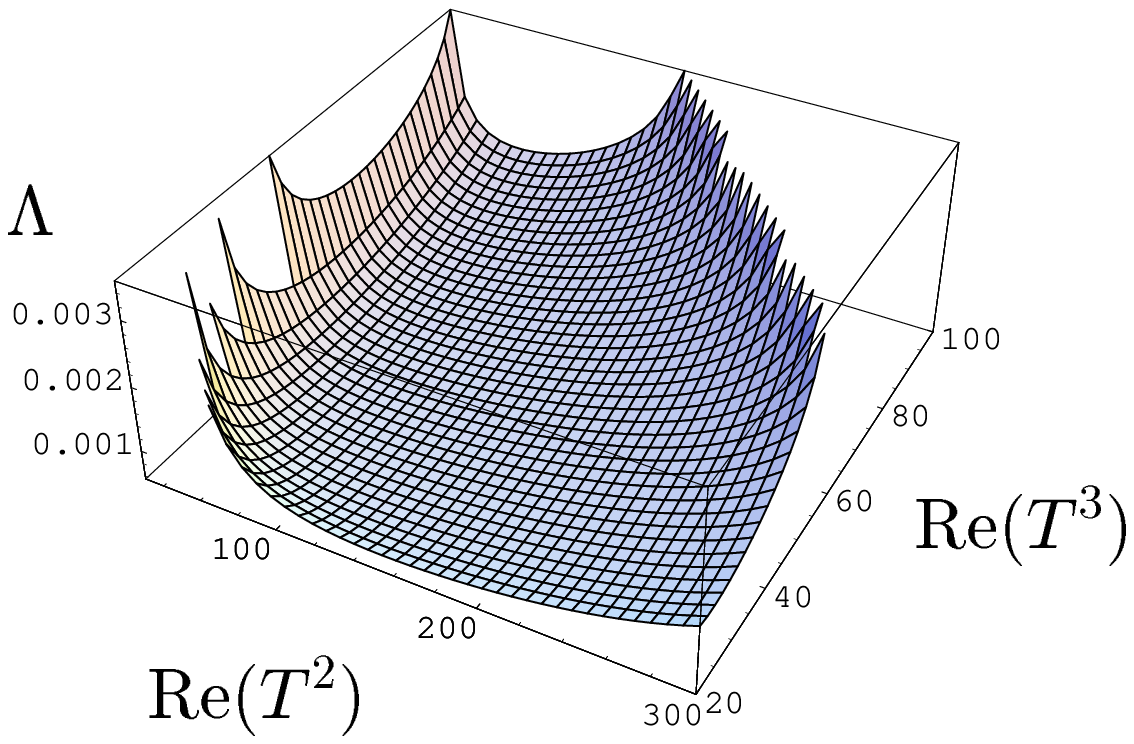, height =11cm}
        \caption{{\rm Model 3}: \, Global minimum of $\Lambda $ as a function of
 $ Re(T^2)$ and $Re(T^3)$. }
        \end{center}
        \end{figure}
\begin{figure}[b]
        \begin{center}  
 \epsfig{file=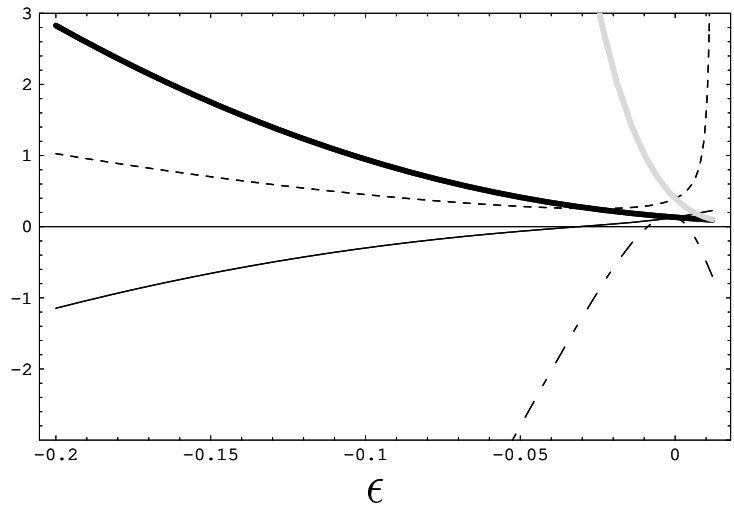, height =9cm}
  \caption{$m^2_{3/2}$ (bold black), $ M^{2 \, \pm}_{1/2}$ (dotted/grey solid curves) and 
$ m^{2 \, \pm}_{ij} $(dashed/black solid curves)
vs  $\epsilon$.}
       \end{center}
       \end{figure}
\begin{figure}[b]
      \begin{center}
\epsfig{file=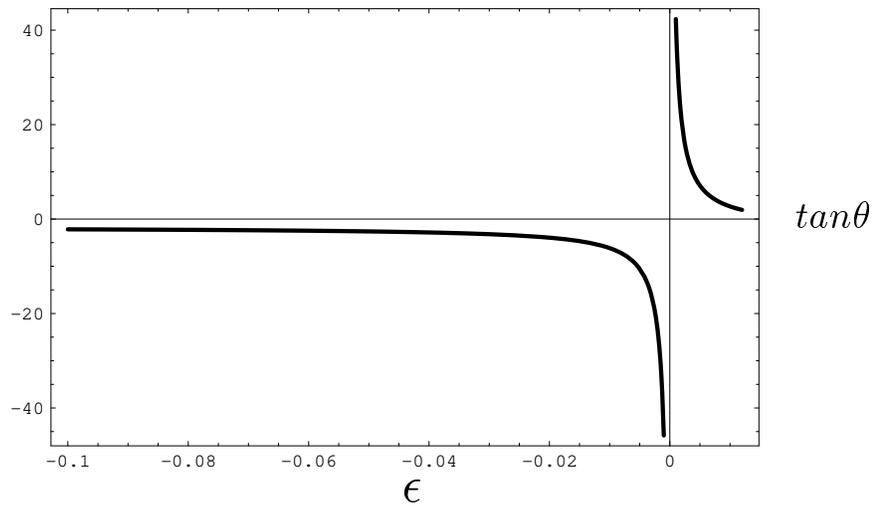, height =9cm}
        \caption{$ \tan( \theta )$ vs $\epsilon$.}
         \end{center}   
\end{figure}     
\begin{figure}[b]
\begin{center}
\epsfig{file=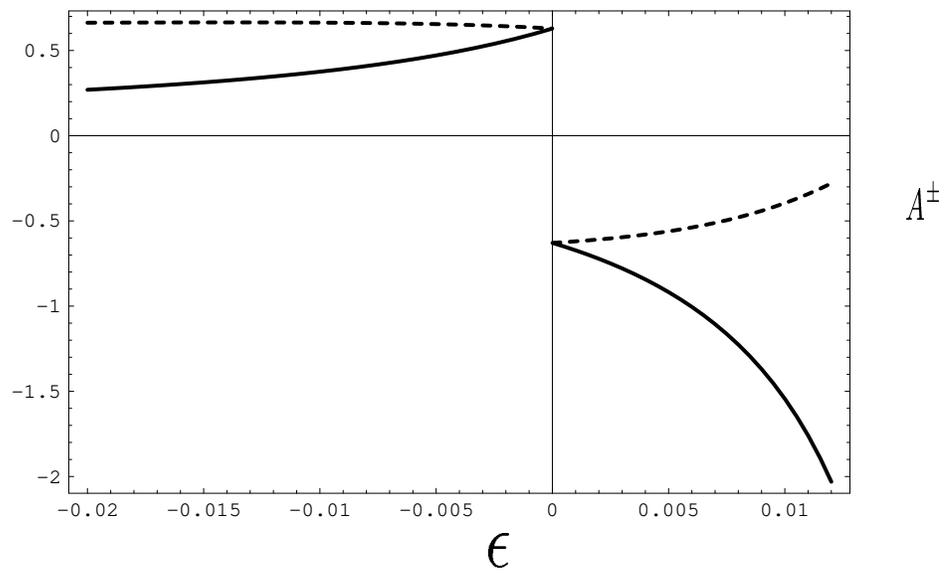, height =10cm}
\caption{ $A^{\pm}$ (dashed/solid curves) vs $\epsilon$.}
 \end{center}   
         \end{figure}
\end{document}